\documentclass[letter,aps,prd,11pt,preprintnumbers,showpacs,showkeys,superscriptaddress,nofootinbib,amsmath,amssymb
]{revtex4-1}
\usepackage{times}
\usepackage{hyperref}
\usepackage{amsmath}
\usepackage{amssymb}
\usepackage{graphicx}
\usepackage{subfigure}
\usepackage{booktabs}
\usepackage{rotating}
\usepackage{longtable}
\usepackage{bm}
\usepackage{float}
\usepackage{epstopdf}
\usepackage{xcolor}
\usepackage{comment}
\usepackage{graphicx}
\usepackage{subfigure}

\usepackage{hyperref}
\hypersetup{
  colorlinks   = true, 	
  urlcolor     = cyan, 	
  linkcolor    = magenta, 	
  citecolor   = blue 	
}
\usepackage{orcidlink}

\makeatletter

\newcommand{\Rmnum}[1]{\expandafter\@slowromancap\romannumeral #1@}
\makeatother

\makeatletter
\newcommand*{\rom}[1]{\expandafter\@slowromancap\romannumeral #1@}
\makeatother
\begin{document}
\title{Accretion Disk Luminosity and Topological Characteristics for a Schwarzschild Black Hole Surrounded by a Hernquist Dark Matter Halo}

\author{Luis M. Nieto\orcidlink{0000-0002-2849-2647}}
\email{luismiguel.nieto.calzada@uva.es}
\affiliation{Departamento de F\'{\i}sica Te\'orica, At\'omica y Optica and Laboratory for Disruptive \\ Interdisciplinary Science (LaDIS), Universidad de Valladolid, 47011 Valladolid, Spain}
\author{Farokhnaz Hosseinifar\orcidlink{0009-0003-7057-451X}}
\email{f.hoseinifar94@gmail.com}
\affiliation{Center for Theoretical Physics, Khazar University, 41 Mehseti Street, Baku, AZ-1096, Azerbaijan}
\author{Kuantay Boshkayev\orcidlink{0000-0002-1385-270X}}
\email{kuantay@mail.ru}
\affiliation{National Nanotechnology Laboratory of Open Type, 050040 Almaty, Kazakhstan}
\affiliation{Al-Farabi Kazakh National University, Al-Farabi ave. 71, 050040 Almaty, Kazakhstan}
\author{Soroush Zare\orcidlink{0000-0003-0748-3386}}
\email{szare@uva.es}
\affiliation{Departamento de F\'{\i}sica Te\'orica, At\'omica y Optica and Laboratory for Disruptive \\ Interdisciplinary Science (LaDIS), Universidad de Valladolid, 47011 Valladolid, Spain}
\author{Hassan  Hassanabadi\orcidlink{0000-0001-7487-6898}}
\email{hha1349@gmail.com}
\affiliation{Departamento de F\'{\i}sica Te\'orica, At\'omica y Optica and Laboratory for Disruptive \\ Interdisciplinary Science (LaDIS), Universidad de Valladolid, 47011 Valladolid, Spain}
\affiliation{Department   of   Physics, Faculty of Science,   University   of   Hradec   Kr\'{a}lov\'{e},  Rokitansk\'{e}ho 62, 500   03   Hradec   Kr\'{a}lov\'{e},   Czechia}

\begin{abstract}

\textbf{Abstract:}
In this work, we study some characteristics and gravitational signatures of the Schwarzschild black hole immersed in a Hernquist dark matter halo (SBH-HDM). We determine the black hole's remnant radius and mass, which provide useful residual information at the end of its evaporation.
We then explore the luminosity of the accretion disk from the SBH-HDM model. In this way, we determine the key orbital parameters of the test particles within the accretion disk, such as angular velocity, angular momentum, energy, and the radius of the innermost stable circular orbit, based on the dark matter model parameters.
We also numerically estimate the accretion disk's efficiency in converting matter into radiation. We also demonstrate that dark matter, which significantly alters the geometry surrounding a Schwarzschild black hole, influences the accretion disk's radiative flux, temperature, differential luminosity, and spectral luminosity.
The stability of a black hole spacetime is determined in the eikonal regime. The Lyapunov exponent is also analyzed to quantify the stability of the particle regime and to demonstrate the infall into or escape from the black hole to infinity, as well as the quasi-normal modes. Finally, some properties of black holes are studied from a topological perspective.

\end{abstract}

\maketitle

\section{Introduction}\label{Sec1}

In recent years, experimental studies in gravitational physics have made considerable progress, providing compelling evidence for the existence of extraordinary astrophysical objects such as black holes \cite{bambi2017black,novikov2013physics,begelman2020gravity,wheeler2007cosmic,wald1992space,novikov2013physics}. In particular, the LIGO/Virgo, Event Horizon Telescope (EHT), and GRAVITY collaborations have achieved groundbreaking discoveries through their coordinated efforts. These include the detection of gravitational wave signals resulting from binary black hole mergers \cite{abbott2016observation,abbott2023gwtc}, observations of the event horizon shadow of the supermassive black holes M87* and Sgr A* within the core of the galaxy M87 \cite{akiyama2019first,akiyama2021first} and Sgr A* \cite{event2022first}, and the detection of infrared flares near our galactic center \cite{baubock2020modeling}. The study of black holes is used to understand the dynamics of galaxies and how they influence their evolution \cite{richstone1998supermassive,heckman2014coevolution,cattaneo2009role,di2008direct,kormendy2013coevolution,morscher2015dynamical}. Black holes exhibit extremely strong gravitational phenomena and high-energy physics, such as quasiperiodic oscillations, the formation of gigantic particle jets, gravitational lenses, and the disruption of nearby stars. From a theoretical point of view, BHs are like a laboratory in which one can test different predictions made by the theories of modified gravity, quantum gravity, and other small- or large-distance corrections to general relativity \cite{bambi2017black}.

In parallel with the study of black holes, dark matter has emerged as a cornerstone of modern cosmology, accounting for a large percentage of the total mass-energy content of the universe \cite{spergel2015dark,clegg2019dark}. The concept of dark matter was first introduced in the early 20th century \cite{bertone2018history,rubin2004brief}. In 1993, while studying the Coma galaxy cluster, Zwicky observed that the visible mass of the galaxies was insufficient to explain the observed gravitational binding, leading him to propose the existence of an additional invisible mass that would provide the gravitational attraction necessary to hold the cluster together \cite{zwicky1933rotverschiebung}. At the end of the 20th century, this idea gained more attention \cite{fall1980formation,white1978core,kapteyn1979first}. For example, the rotation curves of spiral galaxies indicated that stars in the outer regions were moving much faster than expected based on visible mass alone \cite{battaner2000rotation,persic1996universal}. Based on cosmological models, the total mass-energy content of the universe is estimated to be composed of $5\%$ ordinary or baryonic matter, about $27\%$ dark matter, and about $68\%$ dark energy \cite{matarrese2011dark,sciama1993modern,lisanti2017lectures}. The different models in the exploration of dark matter share the common idea that dark matter has mass and contributes to gravity, it does not emit, absorb or reflect light and there are cases in which it has no electromagnetic interactions, although it does interact gravitationally and can affect galaxies, gravitational lenses and cosmic structure \cite{randall2018dark,wechsler2018connection}. Since this type of matter does not interact with electromagnetic forces, it is invisible and its existence is inferred through gravitational effects on visible matter \cite{turner1991dark,bertone2018history,capozziello2012dark}. On the other hand, dark matter behaves like a perfect fluid and can be considered a collisionless fluid, and due to its stability, it has a long lifetime. Dark matter can be classified as cold, warm, or hot, depending on whether the respective masses are of the order $GeV$, $KeV$, or sub-$eV$ \cite{randall2008constraints,trimble1987existence,kunz2016constraints}. This findings led to further investigations into the nature of dark matter. In \cite{konoplya2019shadow}, an analytical expression for the radius of the black hole's shadow was found, assuming a simple spherical configuration of dark matter around it, and their estimates show that dark matter is unlikely to manifest itself in the shadows of galactic black holes, unless its concentration near the black hole is anomalously high. The importance of dark matter goes far beyond its gravitational influence \cite{ghosh2015influences,oks2021brief,sahni20045}, and so, for example, it plays a crucial role in the formation and evolution of cosmic structures, guiding the assembly of galaxies and clusters from primordial density fluctuations imprinted during the inflationary phase of the early Universe \cite{primack1997dark,ratra2008beginning,dayal2018early}. Current cosmological models, particularly the cold dark matter framework, claim that dark matter halos form hierarchically through the merging of smaller structures over time and are made of elementary particles that were not relativistic at early times \cite{zavala2019dark,rubakov2019cosmology,bertone2010particle,short2022dark}.

Dark matter halos provide the gravitational framework within which baryonic matter can cool and condense, ultimately forming stars and galaxies. The dynamics of objects within galaxies are significantly affected by dark matter halos. The structural properties of these haloes, including their density profiles and mass distributions, are characterized using various mathematical models \cite{wechsler2018connection,carr1994baryonic,hogan2000new}. In fact, several density profiles are frequently employed to describe dark matter halos, each one representing the characteristics of different  galaxies. Among them, the Hernquist profile is significant for its effectiveness in describing the internal structure of dark matter halos, characterized by a central peak and a pronounced decrease in density at larger radii \cite{hernquist1990analytical,ghosh2024anisotropic,sadeghian2013dark,figueiredo2023black}.
This profile facilitates the modeling of the gravitational potentials and the understanding of the dynamics of stars within galaxies. 
Recent advances in $N$-body simulations have provided deeper insights into the assembly and evolution of dark matter halos, revealing how their structures influence galaxy formation and the cosmos in general \cite{kamermans2025darkness}.
Furthermore, axial quasi-normal modes (QNM) of black holes immersed in Hernquist-type dark matter halos have been investigated, revealing a universal relationship between the matter environment and redshift QNMs \cite{pezzella2025quasinormal}. Likewise, the distribution of Hernquist-type matter near supermassive black holes has been investigated, focusing on how this distribution affects key orbital properties, such as innermost stable circular orbits (ISCOs), as a function of the surrounding dark matter environment \cite{maeda2025einstein}. Polar perturbations of black holes surrounded by the Hernquist profile and their influence on gravitational wave generation and energy fluxes in extreme mass-ratio spirals \cite{speeney2024black} have also been examined.

As is well known, the investigation of accretion disks surrounding compact objects is a possible way to discern the distinctions between general relativity and several alternative theories of gravity \cite{uniyal2023probing,lambiase2023investigating,khodadi2022probing,allahyari2020magnetically,panotopoulos2021orbits,panah2024accelerating,meng2025bound,wu2023precession,meng2023images,meng2024images,tu2023periodic,zare2024shadows}, as well as the astrophysical environments in which black holes are found \cite{wu2024rotating,pantig2022dehnen,capozziello2023dark,capozziello2023testing,sekhmani2025black,xu2018black}.
Accretion refers to the gravitational infall of matter onto a compact object, such as a black hole, resulting in the redistribution of energy and the generation of high-energy astrophysical phenomena, including relativistic jets and quasars \cite{shakura1973black,novikov1973astrophysics,page1974disk,thorne1974disk,kato2008black}.
Accretion disks are optically thick, geometrically thin structures that are formed as infalling gas conserves angular momentum. 
They initially follow metastable orbits before transitioning to unstable trajectories influenced by the geometry of the spacetime \cite{jusufi2025black}.
 In this way, gravitational energy transforms into thermal radiation through viscous dissipation and magnetohydrodynamic interactions, with emissions reaching their peak in the inner regions of the  disk, where relativistic effects prevail. 
The velocity distribution of accreting matter and the curvature of spacetime are the determining factors of the radiation spectrum, which spans wavelengths from radio to X-rays.
Fundamental geodetic structures, such as photon orbits and the ISCO, define the limits of stable motion in the vicinity of a black hole.
In geometrically thin accretion disks, the ISCO determines the inner radius of the disk, beyond which matter immediately falls into the event horizon.  The boundary is crucial in determining the observable characteristics of the black hole and the accretion efficiency.  Perturbations caused by pressure variations, magnetic fields, or relativistic frame dragging give rise to epicyclic oscillations with distinct radial and vertical components.  In the innermost regions of the accretion disk, these oscillations manifest as the quasi-periodic variability observed in black hole systems and leave a characteristic imprint on the emitted spectrum. Understanding orbital motion and epicyclic frequency is essential for investigating the relativistic accretion dynamics and spacetime geometry of black holes.
Extensive research has been carried out on black hole accretion disks \cite{liu2022thin,harko2009testing,chen2011thin,harko2011thin,johannsen2013inner,mach2013spherical,karkowski2013bondi,yuan2014hot,boshkayev2020accretion,kurmanov2024accretion,kurmanov2025accretion,heydari2021thin,karimov2018accretion,bambi2012code,gyulchev2021image,he2022thin,liu2021thin,jiao2017accretion,murtaza2024evaluation,jiang2024accretion}, attracting significant attention in the scientific literature.

The paper is organized as follows. In section \ref{Sec4}, we present a brief review of the spacetime model of the black hole surrounded by a Hernquist-type dark matter distribution and determine the thermodynamic properties at the horizon. In section \ref{Sec21}, we review the geodesic motion of massive particles traveling through the corresponding static and spherically symmetric spacetime, then we explore the geometrically thin accretion disk framework, inspired by the Novikov-Thorne and Page-Thorne models, and apply it to the SBH-HDM solution. In section \ref{Sec3} we evaluate the impact of the mass density (MD) on the QNMs in the eikonal regime and in section \ref{Sec5} we study topological thermodynamics.
Finally, in section \ref{Sec6} we summarize the main results of this article and establish our conclusions.
The metric signature adopted in this study is $(-+++)$, with the constants set as $G=\hbar=c=1$.

\section{Thermodynamic Properties}\label{Sec4}
\begin{figure}[ht!]
  \includegraphics[width=6.5cm]{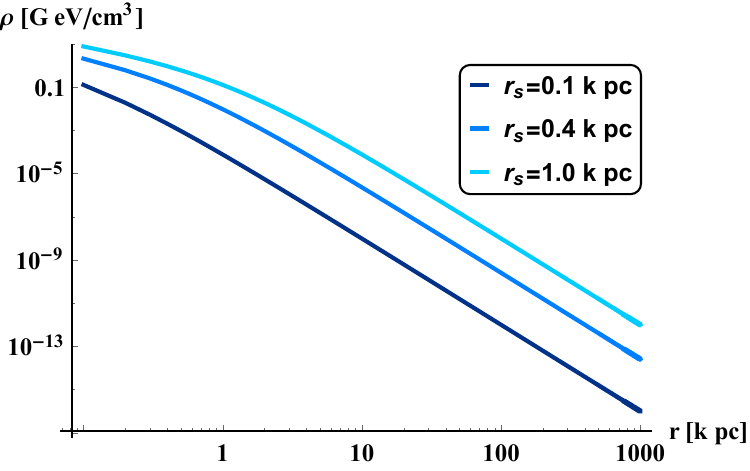} \hspace{-0.2cm}\\
    \caption{Density profile in terms of $r$ considering $\rho_s=1\,G\,eV/cm^3$.}\label{fig:Den}
\end{figure}

\begin{figure}[ht!]
  \includegraphics[width=6.5cm]{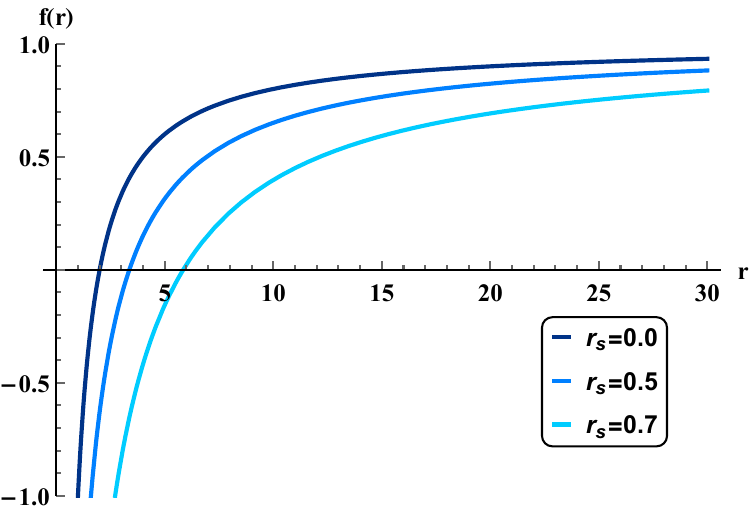} \hspace{-0.2cm}\\
    \caption{Lapse function for $\rho_s=1$ and three values of $r_s$ in terms of $r$.}\label{fig:fr}
\end{figure}

The spacetime metric of pure dark matter can be determined by considering the relationship between the tangential velocity (in the equatorial plane) and the static and spherically symmetric spacetime metric coefficients. Subsequently, by considering the dark matter halo within a general static spherically symmetric spacetime as an extension of the energy-momentum tensors in the Einstein field equation, solutions for spherically symmetric black holes with a dark matter halo will be derived.
In this section, we summarize the main equations characterizing a static, spherically symmetric black hole spacetime surrounded by a region with a Hernquist density profile. For further technical details on the derivation of the spacetime geometry for the SBH-HDM model, which has been used to study the thermodynamic and optical properties of a Schwarzschild black hole immersed in a Hernquist-type dark matter distribution \cite{hernquist1990analytical}, we refer the reader to \cite{jha2025thermodynamics}. Therefore, the SBH-HDM metric is expressed as
\begin{eqnarray}
ds^2= -f(r) dt^2+\frac{dr^2}{g(r)}+h(r) (d\theta^2+\sin^2\theta d\phi^2).\label{ds2}
\end{eqnarray}
The density profile of the Hernquist dark matter halo is given by \cite{mo2010galaxy}
\begin{eqnarray}
\rho(r)=\rho_s\left(\frac{r}{r_s}\right)^{-1}\left(1+\frac{r}{r_s}\right)^{-3},
\end{eqnarray}
where $\rho_s$ refers to the characteristic density and $r_s$ denotes the scale radius of the dark matter halo. Figure \ref{fig:Den} shows the variation of the Hernquist halo density profile for $\rho_s=1$ and three $r_s$ cases.
\\The mass profile of the dark matter distribution is 
\begin{eqnarray}\label{MHalo} 
M_H=\int_{0}^{r}4\pi\rho(r')r'^2dr'=\frac{2 \pi \rho_s r^2 r_s^3}{(r+r_s)^2}.
\end{eqnarray}
Using the tangential velocity of a particle $v_t^2=M_H/r$, considering $g(r)=f(r)$ Eq. \eqref{ds2}, and applying Einstein's field equations, after performing a series of calculations the author of \cite{jha2025thermodynamics} elegantly evaluated the functions $f(r)$ and $h(r)$ that appear in Eq. \eqref{ds2}, which turn out to be
\begin{equation}
f(r)=1 -\frac{2 M}{r}-\frac{4 \pi \rho_s r_s^3}{r+r_s},\qquad
h(r)=r^2, \label{fr}
\end{equation}
where it can be observed that the lapse function $f(r)$ behaves asymptotically flat in $r\to\infty$. Figure \ref{fig:fr} shows the shape of $f(r)$ for three cases of $r_s$.
\\
At the horizon the lapse function is zero, therefore the horizon radius  is obtained as
\begin{eqnarray}\label{horizon}
r_h=\frac{1}{2} \left(2 M-r_s+4 \pi  \rho_s r_s^3+\sqrt{8 M r_s+\left(r_s-2 M-4 \pi  \rho_s r_s^3\right)^2}\right),
\end{eqnarray}
and it is evident that in the limit of $r_s\to 0$, the horizon radius of the black hole of Eq. \eqref{ds2} tends to the Schwarzschild black hole horizon.

The mass of the black hole as a function of horizon radius is obtained from $f(r_h)=0$ and is
\begin{eqnarray}\label{Mh}
M(r_h)=r_h \left(\frac{1}{2}-\frac{2 \pi\rho_s r_s^3}{r_h+r_s}\right),
\end{eqnarray}
which approaches the mass of the Schwarzschild black hole in the limit of $r_s\to 0$ or $\rho_s\to 0$.

\begin{figure}[ht!]
  \includegraphics[width=6.5cm]{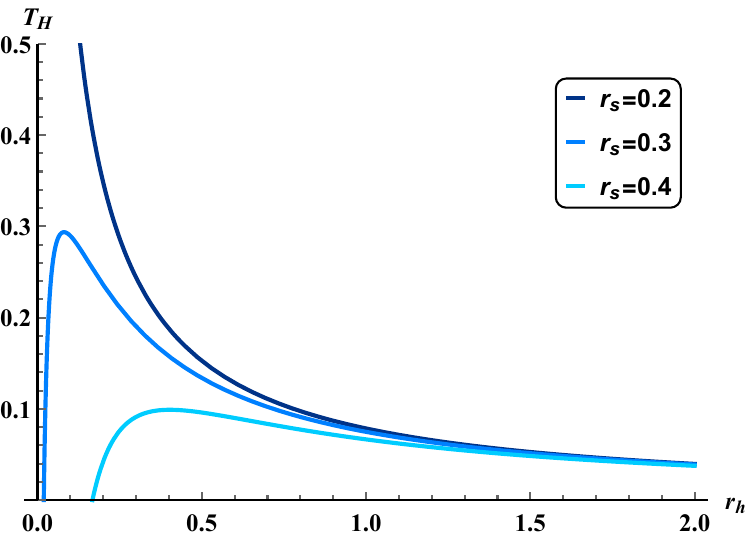} \hspace{-0.2cm}\\
    \caption{Hawking temperature curves for $\rho_s=1$ and three values of $r_s$ versus the horizon radius.}\label{fig:TH}
\end{figure}

The Hawking temperature is calculated from $1/(4\pi)\,\partial_r f(r)\big|_{r=r_h}$ and by substituting the mass given in Eq.  \eqref{Mh}, the following result is obtained \cite{hawking1974black,hawking1983thermodynamics,fredenhagen1990derivation}
\begin{eqnarray}\label{TH}
T_H=\frac{1}{4 \pi  r_h}-\frac{\rho_s r_s^4}{r_h (r_h+r_s)^2},
\end{eqnarray}
where it is evident that in the limit of $r_s\to 0$ or $\rho_s\to 0$, the equation \eqref{TH} tends to the Hawking temperature of the Schwarzschild black hole. Figure \ref{fig:TH} illustrates the behavior of the Hawking temperature in terms of the horizon radius for three values of $r_s$.

\begin{figure}[ht!]
  \includegraphics[width=6.5cm]{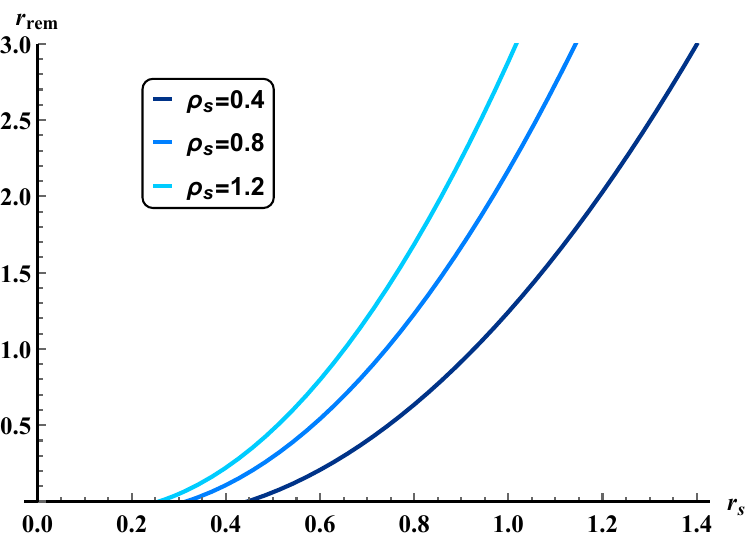} \hspace{-0.2cm}\\
    \caption{Remnant radius of the black hole versus $r_s$ for three different values of $\rho_s$.}\label{fig:rRem}
\end{figure}

The remnant radius of the black hole, which indicates the radius after the evaporation process, is calculated from $T_H\big|_{r=r_{\text{rem}}}=0$ and reads \cite{delhom2019absorption,giddings1992black,dong2025some}
\begin{eqnarray}\label{rem}
r_{\text{rem}}=r_s \left(2r_s \sqrt{\pi \rho_s} -1\right).
\end{eqnarray}
Obviously it can be concluded that maintaining the condition $2 \sqrt{\pi \rho_s} r_s>1$, the remnant radius is positive.
Figure \ref{fig:rRem} illustrates the variation of the remnant radius in terms of $r_s$ for three values of $\rho_s$.
As expected from Eq. \eqref{rem} and obvious in figure \ref{fig:rRem}, increasing $r_s$ and $\rho_s$ results in an increase in the remnant radius. Furthermore, if the horizon radius in the black hole mass Eq. \eqref{Mh} is substituted by the remnant radius from Eq. \eqref{rem}, the remnant mass of the black hole can be obtained.

\section{Radiative Flux and Spectral Luminosity of the
Accretion Disk}\label{Sec21}

\begin{figure}[ht!]
\centering
  \includegraphics[width=6.5cm]{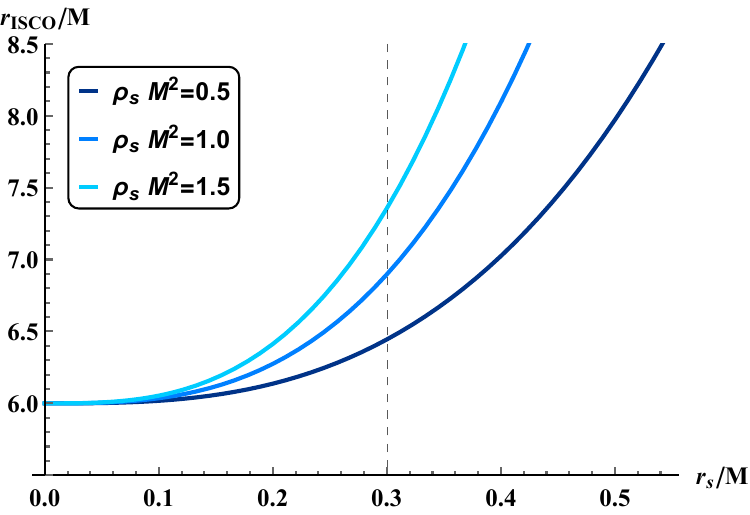} \hspace{1.2cm}
  \includegraphics[width=6.5cm]{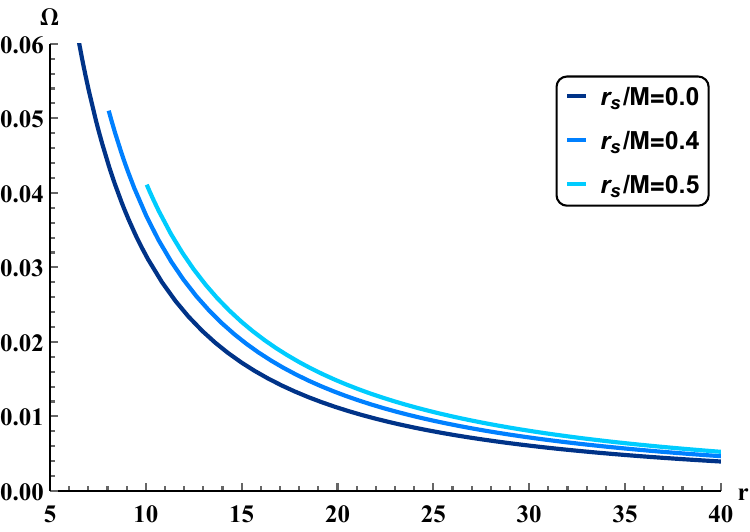} \\
  \includegraphics[width=6.5cm]{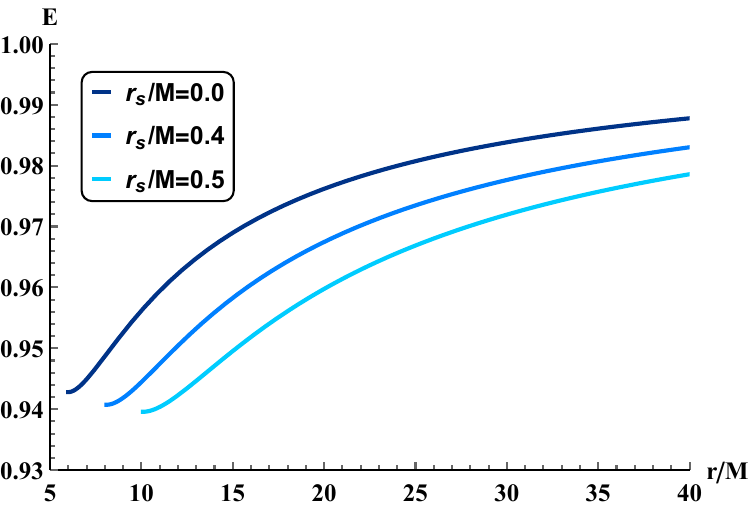} \hspace{1.2cm}
  \includegraphics[width=6.5cm]{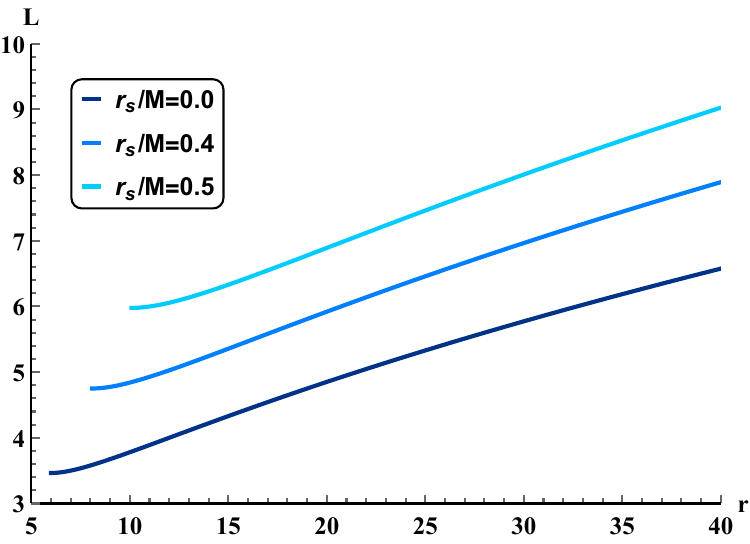} \\
    \caption{The radius of the ISCO, $r_{\text{ISCO}}$, in terms of $r_s/M$, and the behavior of angular velocity, angular momentum, and energy considering $\rho_s M^2=1$.}\label{fig:AC1}
\end{figure}

We begin this section with a brief review of the physical characteristics of thin accretion disks according to the Novikov-Thorne model \cite{novikov1973astrophysics}, which is a more general version of the Shakura-Sonyaev model \cite{shakura1973black}.
The model is based on some conventional assumptions: 
I) the spacetime of the central massive entity is stationary, axisymmetric, and asymptotically flat; 
II) the disk mass has no influence on the background metric; 
III) the accretion disk is assumed to be geometrically thin, with its vertical scale height essentially negligible compared to its radial extent; IV) the orbiting particles are confined between the outer radius $r_{\mathrm{OUT}}$ and the innermost stable circular orbit $r_{\mathrm{ISCO}}$, which defines the inner boundary of the disk; 
V) the accretion disk is confined to the equatorial plane $(\theta = \pi/2)$ of the central compact object, so that its surface lies perpendicular to the black hole's spin axis;
VI) under the assumption of hydrodynamic and thermodynamic equilibrium, the disk emits electromagnetic radiation with a blackbody spectrum characteristic of its local temperature; 
VII) the mass accretion rate, $\dot{m}$, is considered constant in time \cite{he2022thin}. 
The radiation emanating from the accretion disk can be strongly affected by the parameters of the black hole, making this research essential for understanding the physical characteristics of black holes and their effects on the results of astrophysical observations.
In this work, we first investigate the ISCO of massive test particles within the black hole's gravitational field, described by the spacetime metric \eqref{ds2}. 
Next, we explore the radiative properties of the disk, which are essential for characterizing its observable signatures.
To do this, it is first necessary to explore the geometry of the space where particles travel near a compact object \cite{harko2011thin}. The Lagrangian governing the geodesic motion of test particles moving around a massive object is \cite{berezhiani2012black,gibbons2015jacobi,bader2005quantum}:
\begin{equation}\label{Lag}
2\mathcal{L} = g_{\mu\nu}\dot{x}^{\mu}\dot{x}^\nu= -f(r)\dot{t}^2+\frac{\dot{r}^2}{f(r)}+h(r)(\dot{\theta}^2+\sin^2\theta\dot{\phi}^2),
\end{equation}
where the dot represents the derivative with respect to $\tau$, which is an affine parameter along the geodesic $x^{\mu}(\tau)$. The metric coefficients are only functions of the radial coordinate $r$.  Furthermore, we employ an equatorial approximation, indicating that $|\theta - \pi/2| << 1$. Since the aforementioned metric does not depend on the coordinates $t$ and $\phi$, we expect to have two constants of motion, namely the energy $E$ and the angular momentum per unit rest-mass $L$ \cite{perlick2022calculating,ono2019effects,bakopoulos2024exact}, given by 
\begin{eqnarray}\label{EL}
E=f(r)\dot{t},\qquad L=h(r)\dot{\phi},
\end{eqnarray}
A significant attribute of the thin accretion disk is the effective potential $V_{\text{eff}}(r)$. Therefore, the radial equation of motion is derived from the relation $\mathcal{L} = -1/2$, considering Eqs.~ \eqref{Lag} and \eqref{EL}, as follows:
\begin{eqnarray}\label{traj}
\frac{1}{2}\dot{r}^2+V_{\text{eff}}(r)=\frac{1}{2}E^2,
\end{eqnarray}
in which the effective potential is
\begin{equation}\label{veff}
V_{\text{eff}}(r)=\frac{f(r)}{2}\left(1+\frac{L^2}{r^2}\right).
\end{equation}
The circular orbit moves through the minimum point of the effective potential. In the absence of a minimum (i.e., when the potential exhibits``smooth" behavior), circular orbits are unstable at a specific point.
The ISCO is related to a marginally stable circular orbit. For such a trajectory, the conditions $\frac{dr}{d\tau} = 0$ and $\frac{d^{2} r}{d\tau^{2}} = 0$ must be met. To ensure marginal stability, the additional requirement ${d^{3} r}/{d\tau^{3}} = 0$ must also be met. Applying these requirements to  Eq. \eqref{traj}, the following equivalent set of expressions is obtained \cite{jiang2024accretion}:
\begin{eqnarray}\label{Veff}
V_{\text{eff}}(r)=\frac{1}{2}E^2,\qquad \partial_r V_{\text{eff}}(r)=0, \qquad\text{and}\qquad \partial^2_r V_{\text{eff}}(r)=0.
\end{eqnarray}
All circular orbits beyond the ISCO are stable, unless there is an outermost stable circular orbit. Only the first two equations of \eqref{Veff} are required for general circular orbits. These equations can be used to obtain both the specific energy $E$ and angular momentum $L$ of a massive particle in a circular orbit of radius $r$ \cite{kurmanov2022accretion}:
\begin{eqnarray}
\label{Lr}
&&L(r)=\frac{\Omega(r)h(r)}{\sqrt{f(r)+\Omega(r)^2 h(r)}}
=\frac{r \sqrt{M (r+r_s)^2+2 \pi  \rho_s r^2 r_s^3}}{\sqrt{r \left((r+r_s)^2-2 \pi  \rho_s r_s^3 (3 r+2 r_s)\right)-3 M (r+r_s)^2}},\\ [1ex]
\label{Er}
&&E(r)=\frac{f(r)}{\sqrt{f(r)+\Omega(r)^2 h(r)}}=\frac{(r-2 M) (r+r_s)-4 \pi  \rho_s r r_s^3}{\sqrt{r \left((r-3 M) (r+r_s)^2-2 \pi  \rho_s r r_s^3 (3 r+2 r_s)\right)}}.
\end{eqnarray}
The angular velocity of the particle in the orbit is given by
\begin{align}
\Omega(r)&=\sqrt{\frac{\partial_r f(r)}{\partial_r h(r)}}=\frac{\sqrt{M (r+r_s)^2+2 \pi  \rho_s r^2 r_s^3}}{r^{3/2} (r+r_s)},
\end{align}
where Eqs. \eqref{Lr}, \eqref{Er} and \eqref{EL} have been used \cite{novikov1973astrophysics,kurmanov2022accretion}. The requirement $\partial^2_r V_{\text{eff}}(r)=0$ gives $r_{\rm ISCO}$.  Hence, the radius of the ISCO surrounding the SBH-HDM can be determined by  \cite{collodel2021circular}
\begin{equation}\label{IS1}
E(r)^2\partial^2_r h(r)-L(r)^2\partial^2_r f(r)-\partial^2_r(f(r) h(r))=0.
\end{equation}
By employing Eqs. \eqref{fr}, \eqref{Lr}, and \eqref{Er}, Eq. \eqref{IS1} can be expressed in the following form:
\begin{equation}\label{IS2}
\!\!\!\!\! \left. \frac{2 M (6 M-r) (r+r_s)^3-4 \pi  \rho_s r r_s^3 \left(r^2 (r+3 r_s)-2 M \left(6 r^2+9 r r_s+r_s^2\right)\right)+48 \pi ^2 \rho_s^2 r^3 r_s^6}{r (r+r_s) \left((r-3 M) (r+r_s)^2-2 \pi  \rho_s r r_s^3 (3 r+2 r_s)\right)}\right|_{r_{\text{ISCO}}}\!\!\!\!=0.
\end{equation}
Figure \ref{fig:AC1} illustrates the variation of the ISCO radius as a function of $r_s/M$, together with the  angular velocity, angular momentum, and energy profiles for $\rho_s M^2=1$ and selected values of $r_s/M$, plotted as functions of the radial coordinate $r/M$.
As expected, in the limit of $r_s\to 0$, the black hole metric function in Eq. \eqref{fr} tends to that of the Schwarzschild black hole.
Consequently, the initial value of the black hole's ISCO radius is similar to the Schwarzschild black hole's ISCO radius, which is $6M$.
An increase in either the $r_s/M$ or $\rho_s M^2$ parameters leads to an increase in the size of the ISCO radius.
Furthermore, for small $r_s/M$, the sensitivity of $r_{\text{ISCO}}$ to variations in $\rho_s$ is low. However, as $r_s$ increases, the sensitivity of $r_{\text{ISCO}}$ to changes in $\rho_s$ becomes more pronounced.
On the other hand, at constant $\rho_s M^2$, the sensitivity of $\Omega(r)$ to changes in $r_s/M$ is low, while the sensitivities of $L(r)$ and $E(r)$ are higher. Similarly, in the presence of Hernquist dark matter, particles on the accretion disk exhibit higher values of velocity and angular momentum, as well as lower energy compared to the Schwarzschild black hole. As $r_s/M$ increases, the disparity in these quantities for the Hernquist dark matter black hole compared to the Schwarzschild black hole becomes more pronounced.

The energy flux emitted by the accretion disk surrounding the black hole refers to the amount of energy radiated per unit area per unit time, and is measured at the ISCO. It is obtained from \cite{Noble2009Direct,McClintock2011Measuring,Zuluaga2021Accretion}
\begin{eqnarray}
\mathcal{F}(r)=-\frac{\dot{m}}{4\pi\sqrt{h(r)}}\frac{\partial_r \Omega(r)}{(E(r)-\Omega(r) L(r))^2}\int_{r_{\text{ISCO}}}^{r}(E(R)-\Omega(R) L(R))\partial_{R}L(R) d R,
\end{eqnarray}
where $\dot{m}$ is the accretion rate mass.
For simplicity, we can assume it is constant and take $\dot{m} = 1$, which is equivalent to taking into account the normalized flux per unit accretion rate, i.e., $\mathcal{F}(r)/\dot{m}$ \cite{kurmanov2022accretion}. 
Using the  Stefan--Boltzmann law, $\mathcal{F}(r)=\sigma_{SB}T(r)^4$, where $T(r)$ refers to the radiation temperature and $\sigma_{SB}$ indicates the Stefan--Boltzmann constant.

\begin{figure}[ht!]
  \includegraphics[width=6.5cm]{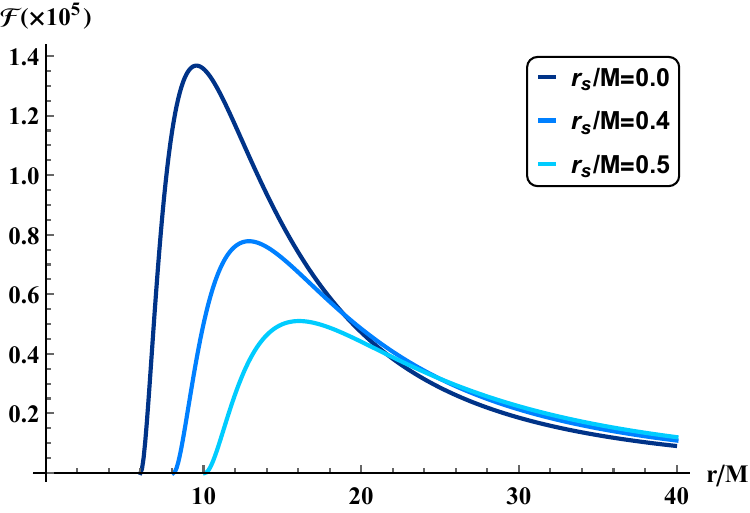} \hspace{1.2cm}
  \includegraphics[width=6.5cm]{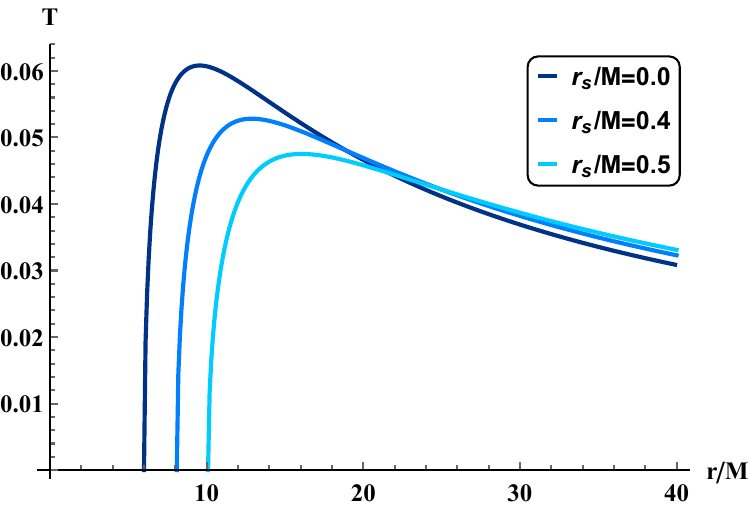}
    \caption{For $\rho_s M^2=1$, the energy flux is represented in the left panel and the radiation temperature in the right panel.}\label{fig:AC2}
\end{figure}

Figure \ref{fig:AC2} shows the energy flux and radiation temperature curves considering $\rho_s M^2=1$ for three different values of $r_s/M$.
It is observed that as $r/M$ increases, the energy flux reaches a maximum value before subsequently decreasing, and as $r_s/M$ increases, the maximum value of $\mathcal{F}(r)$, which is located at $\partial_r \mathcal{F}(r)\big|_{r=r_c}=0$, also decreases, occurring at a smaller $r/M$ value.
Furthermore, although the energy flux for the Schwarzschild black hole is initially higher at small radii compared to the black hole with Hernquist dark matter, the magnitude of energy flux increases in the presence of Hernquist dark matter at larger radii. The larger the $r_s/M$, the greater the value of $\mathcal{F}$. A similar trend is observed for the radiation temperature.

\begin{figure}[ht!]
  \includegraphics[width=5.3cm]{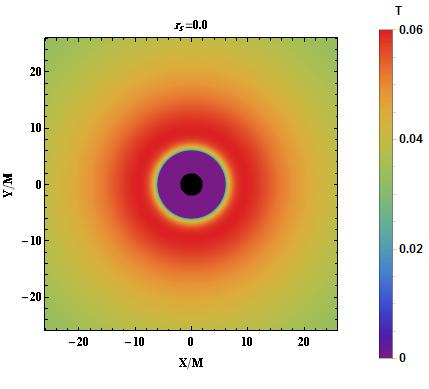} \hspace{0.1cm}
  \includegraphics[width=5.3cm]{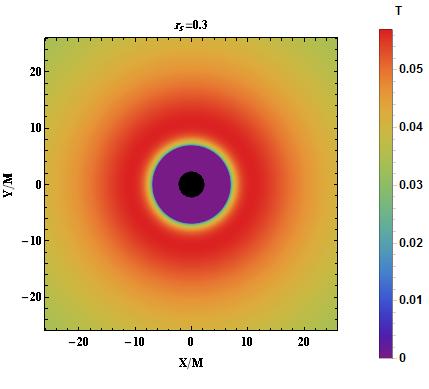} \hspace{0.1cm}
  \includegraphics[width=5.3cm]{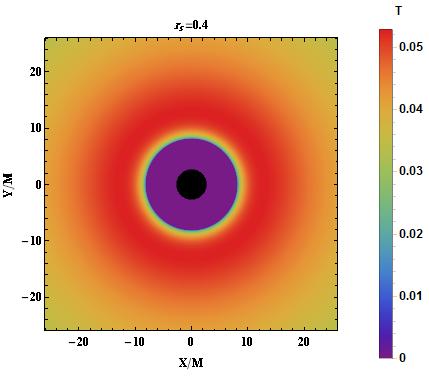} \hspace{-0.1cm}
    \caption{Radiation temperature for $\rho_s M^2=1$ in the equatorial Cartesian coordinate plane $x-y$.}\label{fig:AC4}
\end{figure}

\begin{figure}[ht!]
  \includegraphics[width=5.3cm]{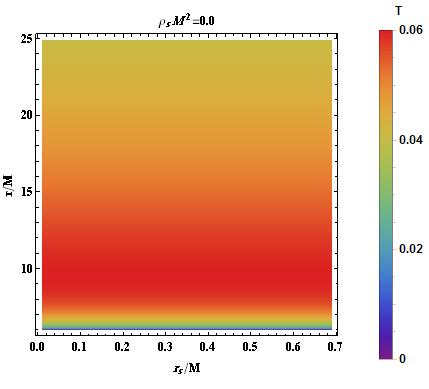} \hspace{0.1cm}
  \includegraphics[width=5.3cm]{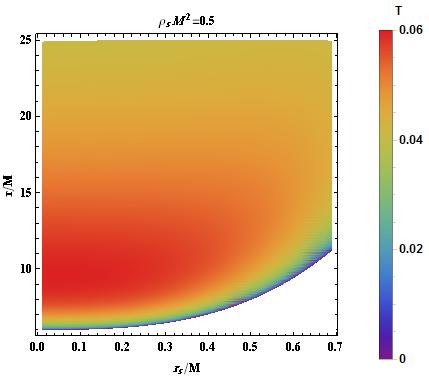} \hspace{0.1cm}
  \includegraphics[width=5.3cm]{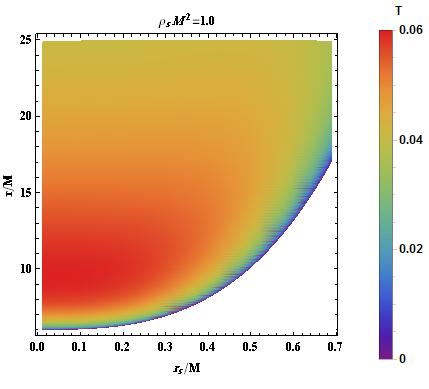} \hspace{-0.1cm}
    \caption{The behavior of the radiation temperature when varying $r_s/M$ and $r/M$ for three options of $\rho_s M^2$.}\label{fig:AC6}
\end{figure}

Figures \ref{fig:AC4} and \ref{fig:AC6} depict the behavior of the radiation temperature density. In figure \ref{fig:AC4}, considering $\rho_s M^2=1$, the radiation temperature variations are graphed on an $x-y$ Cartesian plane corresponding to the equator.
It is evident that, after crossing the horizon (marked by the black disk), the emitted temperature is initially zero, reaches its maximum after the ISCO radius, and then gradually decreases. Furthermore, as anticipated in figure \ref{fig:AC2}, while increasing the parameter $r_s/M$ leads to a reduction in the maximum source radiation temperature, this maximum occurs at a larger radius. Furthermore, while at smaller radii, a lower $r_s/M$ corresponds to higher temperatures, at larger radii, a higher $r_s/M$ results in higher temperature values.

Figure \ref{fig:AC6} shows the temperature behavior in terms of $r/M$ and $r_s/M$. As indicated in figure \ref{fig:AC2}, as $r_s/M$ increases, the maximum radiation temperature decreases and occurs at a larger radius.

By fixing $\dot{m}$, the differential luminosity at infinity, which indicates the total amount of energy emitted by the accretion disk per unit time that reaches an observer at infinity, is determined as \cite{Boshkayev2021Luminosity,boshkayev2020accretion,boshkayev2024luminosity}
\begin{eqnarray}
\frac{d \mathcal{L}_\infty}{d \ln r}=4\pi r^2 E(r) \mathcal{F}(r).
\end{eqnarray}
The spectral luminosity distribution at spatial infinity is calculated from \cite{d2023constraining,boshkayev2022accretion}
\begin{equation}
\nu\mathcal{L}_{\nu,\infty}=\frac{15}{\pi^4}\int_{r_i}^\infty \left(\frac{d \mathcal{L}_\infty}{d \ln r}\right)\frac{\left(u(r)y\right)^4}{M^2_T \mathcal{F}(r)}\frac{1}{\exp\left(\dfrac{u(r)y}{\left[M^2_T \mathcal{F}(r)\right]^{1/4}}\right)-1}d\ln r,
\end{equation}
where 
\begin{equation}
u(r)=\frac{1}{\sqrt{f(r)+\Omega^2(r)h(r)}}.
\end{equation}
Here $M_T$ is the total mass of the black hole and the halo, with the halo mass given in Eq. \eqref{MHalo}, and the variable $y$ is $y=\hbar\nu/k T_*$, where $\hbar$ is Plank's constant, $\nu$ is the radiation frequency, $k$ indicates the Boltzmann constant and $T_*=\dot{m}/\left(4\pi M^2_T\sigma_{\rm SB}\right)$.

\begin{figure}[ht!]
  \includegraphics[width=6.5cm]{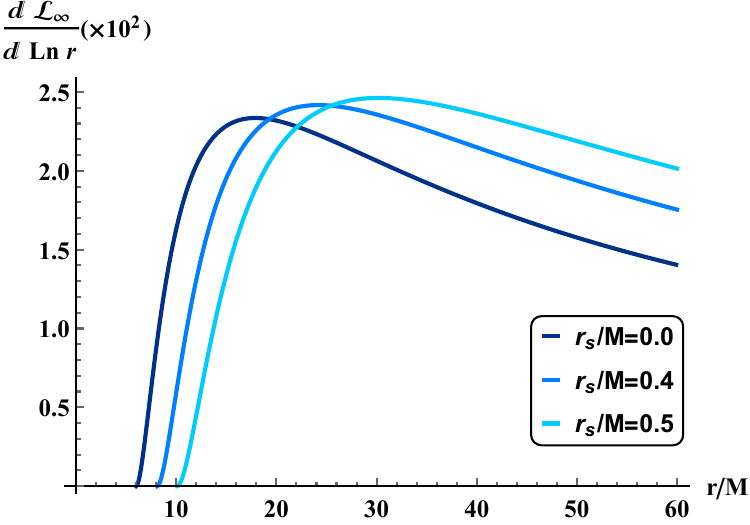} \hspace{0.5cm}
  \includegraphics[width=7cm]{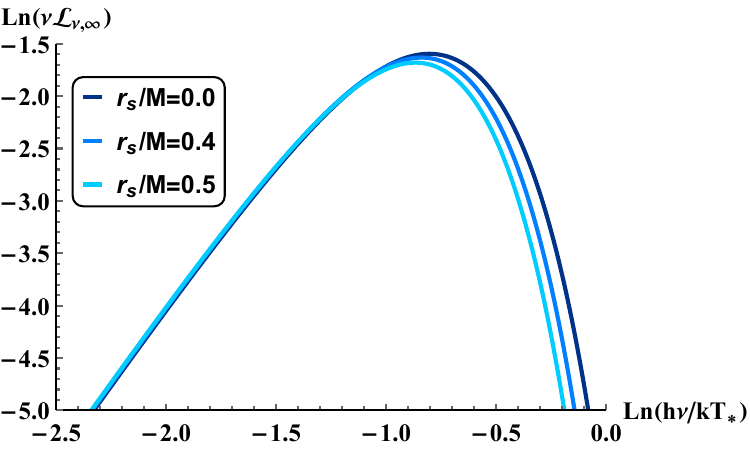} \hspace{-0.2cm}
    \caption{For $\rho_s M^2=1$, the differential luminosity at infinity is shown in the left panel and the observed spectral luminosity in the right panel.}
    \label{fig:AC3}
\end{figure}

Figure \ref{fig:AC3} represents the variation of the differential luminosity at infinity and the spectral luminosity for $\rho_s M^2=1$ and three values of $r_s/M$.
It is observed that the differential luminosity at infinity exhibits energy flux-like behavior, and at constant $\rho_s M^2$, an increase in $r_s/M$ results in a lower maximum value of $\frac{d \mathcal{L}_\infty}{d \ln r}$, which occurs at lower $r/M$. Furthermore, it is evident that while at small $r/M$ the spectral luminosity of the black hole in the presence of Hernquist dark matter is higher and increases with higher $r_s/M$, at large $r/M$ the trend is reversed, and the Schwarzschild black hole exhibits a higher luminosity.

\begin{figure}[ht!]
	\centering
  \includegraphics[width=6.5cm]{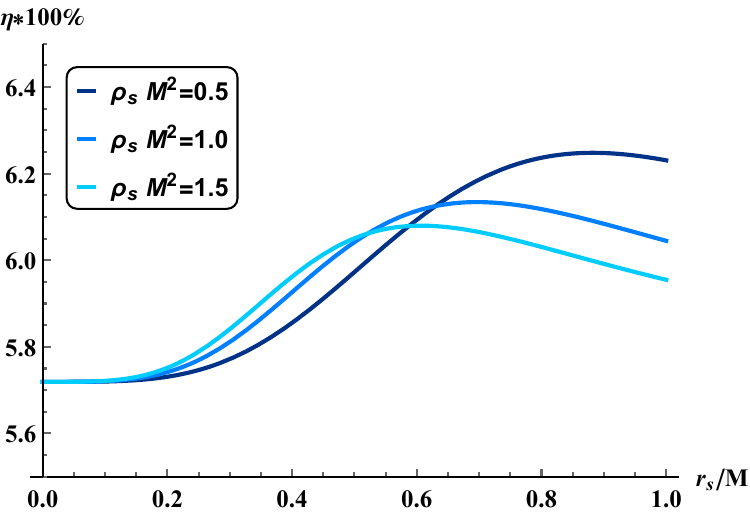} \hspace{-0.2cm}
    \caption{Radiative efficiency in terms of $r_s/M$. For $r_s/M=0$, the magnitude of the efficiency is $\eta= 5.7191\%$, which is the Schwarzschild black hole efficiency. }\label{fig:AC5}
\end{figure}

Compared to a Schwarzschild black hole in vacuum, it is also intriguing to determine the radiative efficiency of the source, which is the amount of energy from the disk's rest mass that is transformed into radiation, which is
\begin{eqnarray}
\eta=\frac{\mathcal{L}_{\rm{bol}}}{\dot{m}}\simeq 1-E(r_{\rm ISCO}),
\end{eqnarray}
where, in light from a Schwarzschild black hole, one obtains the standard result $\eta \approx 5.72\%$ \cite{kurmanov2022accretion}. Here, the bolometric luminosity of the accretion disk is denoted by $\mathcal{L}_{\rm{bol}}$.
Figure \ref{fig:AC5} shows the variations in efficiency in the presence of Hernquist dark matter as the parameter $r_s/M$ increases.
As can be seen, at $r_s/M \to 0$, the efficiency equals that of the Schwarzschild black hole, and as the quantity $r_s/M$ increases, this improves. It is also observed that, while for a small value of $r_s/M$, the efficiency increases with increasing $\rho_s M^2$, as $r_s/M$ gradually increases, the magnitude of this quantity becomes larger for smaller values of $\rho_s M^2$.

\section{Quasi-normal Modes}\label{Sec3}

Quasi-normal modes (QNMs) correspond to damped oscillations defined by complex frequencies that encode energy dissipation and reveal key physical properties of black holes, such as mass, charge, and angular momentum. In this section, we explore scalar field perturbations in SBH-HDM spacetime and derive the corresponding effective potential governing their dynamics.

\begin{figure}[ht!]
  \includegraphics[width=6.5cm]{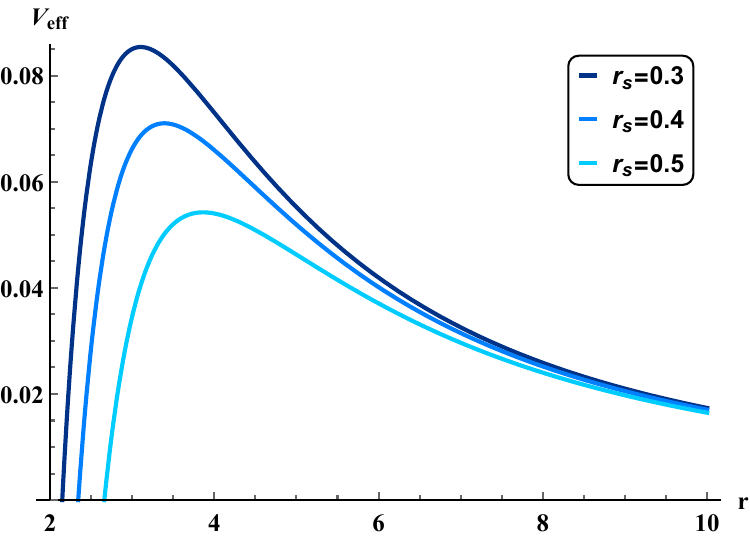} \hspace{0.5cm}
  \includegraphics[width=6.5cm]{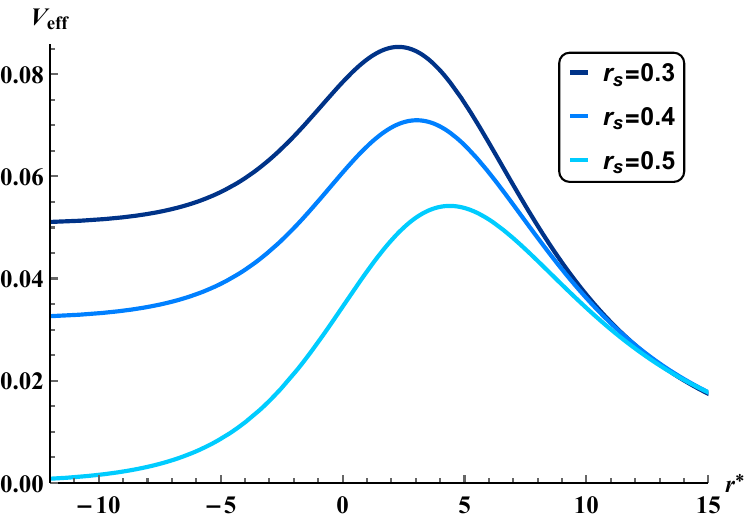} \hspace{-0.2cm}
    \caption{Effective potential curves considering $M=1$, $\rho_s=0.5$ and $l=2$. In the left panel, as a function of $r$, and in the right panel, varying $r^*$.}\label{fig:Veff}
\end{figure}

We consider now a massless scalar field in a curved spacetime whose Klein-Gordon equation has the following form \cite{landau2013classical,lefloch2016global}
\begin{align}\label{WaveF}
\frac{1}{\sqrt{-g}}\partial_\mu(\sqrt{-g}g^{\mu\nu}\partial_\nu\Psi_{\omega lm}(\textbf{r},t))=0,
\end{align}
where $g_{\mu\nu}$ is the metric element of Eq. \eqref{ds2}, $g$ is the determinant, and $g^{\mu\nu}$ is the inverse of $g_{\mu\nu}$. We assume that the wavefunction is given by \cite{konoplya2011quasinormal}
\begin{align}\label{psi}
\Psi_{\omega lm}(\textbf{r},t)=\frac{R_{\omega l}(r)}{r}Y_{lm}(\theta,\,\phi)e^{-i\omega t},
\end{align}
where $Y_{lm}(\theta,\,\phi)$ is the spherical harmonics and $\omega$ is the frequency.
Here, as usual, $l$ represents the angular quantum number (multipole) and $m$ denotes integers, provided that $l > |m|$.
By inserting Eq. \eqref{psi} into Eq. \eqref{WaveF} and separating the angular variables, we can obtain a Schr\"{o}dinger-like equation:
\begin{align}\label{psir}
f(r)\frac{d}{dr}\left(f(r)\frac{dR_{\omega l }(r)}{dr}\right)+(\omega^2-V(r))R_{\omega l}(r)=0.
\end{align}
Considering the tortoise coordinates as $dr^*=dr/f(r)$ \cite{thorne2000gravitation}, Eq. \eqref{psir} can be written as \cite{konoplya2002quasinormal}
\begin{align}\label{r_rstar}
\frac{d^2}{dr^{*^2}}R_{\omega l}(r^*)+(\omega^2-V_{\rm eff}(r^*))R_{\omega l}(r^*)=0,
\end{align}
where $V_{\rm eff}$ refers to the Regge-Wheeler effective potential, which is used to examine the QNM of black holes, and for scalar fields it is defined in Ref. \cite{regge1957stability}:
\begin{eqnarray}
V_{\rm eff}(r)=f(r)\left(\frac{1}{r}\frac{\partial f(r)}{\partial r}+\frac{l (l+1)}{r^2}\right).\label{VRW}
\end{eqnarray}
We hypothesize that, when $r_{s}$ or $\rho_{s}$ are held constant, increasing the other parameter at small scales significantly alters the height of the effective potential barrier. This indicates that the presence of dark matter halos can have a substantial impact on the geometry of the spacetime surrounding Schwarzschild black holes.
Figure \ref{fig:Veff} illustrates the variation of the effective potential in terms of $r$ and $r^*$.
As shown, increasing the parameter $r_s$ leads to a decrease in the magnitude of the effective potential, resulting in its maximum occurring at a larger radius.
Furthermore, the correlation between the maximum of $V_{\rm eff}$ and the quasi-normal frequencies implies that changes in $r_{s}$ or $\rho_{s}$ could have comparable impacts on the quasi-normal frequencies.

Considering $Q(r^*)=\omega^2-V_{\rm eff}(r^*)$, we obtain the following differential equation \cite{heidari2023investigation}
\begin{eqnarray}
\frac{d^2 R_{\omega l}(r^*)}{dr^{*^2}}+ Q(r^*)R_{\omega l}(r^*)=0.
\end{eqnarray}
In the eikonal limit, the term $\dfrac{f(r)}{r}\dfrac{\partial f(r)}{\partial r}$ is negligible, also as $l\gg 1$, so $l(l+1)\approx l^2$ and $Q$ takes the following form
\begin{eqnarray}\label{Qr}
Q=\omega^2-\frac{f(r)}{r^2}l^2.
\end{eqnarray}
Assuming that $Q$ has a maximum at $\partial_r Q\big|_{r=r_c}=0$, then Eq. \eqref{Qr} can be written as
\begin{eqnarray}
Q\equiv Q(r_c)+\underbrace{\frac{1}{1 !}\partial_r Q\big|_{r=r_c}(r-r_c)}_0+\frac{1}{2 !}\partial^2_r Q\big|_{r=r_c}(r-r_c)^2+O(r^3),
\end{eqnarray}
where $Q(r_c)=\omega^2-\dfrac{f(r_c)}{r_c^2}l^2$. 
Furthermore, the maximum of $Q$ is calculated from $r_c \partial_r f(r)\big|_{r=r_c}-2f(r_c)=0$, which is similar to the circular orbit or circular null geodesic \cite{virbhadra2000schwarzschild,virbhadra2002gravitational}. 
We therefore conclude that there exists a well-defined geometrical-optical limit, called the eikonal limit, applicable to a wide class of massless perturbations \cite{cardoso2009geodesic}.

\begin{figure}[ht!]
  \includegraphics[width=6.5cm]{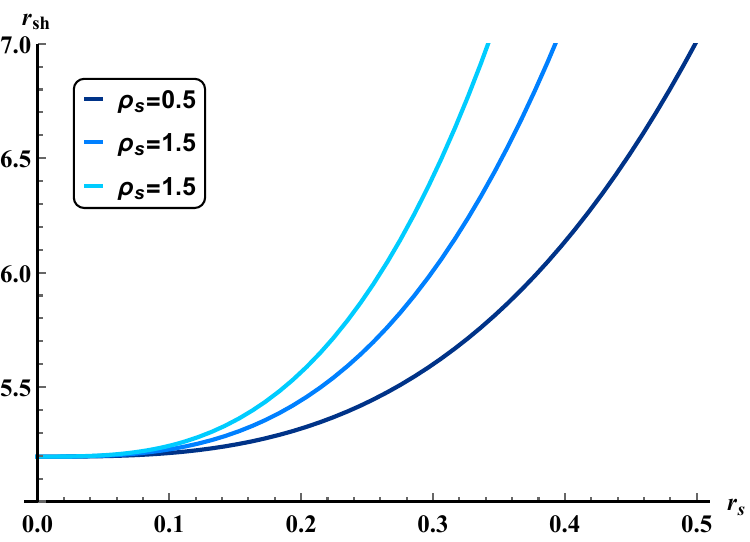} \hspace{0.5cm}
  \includegraphics[width=6.5cm]{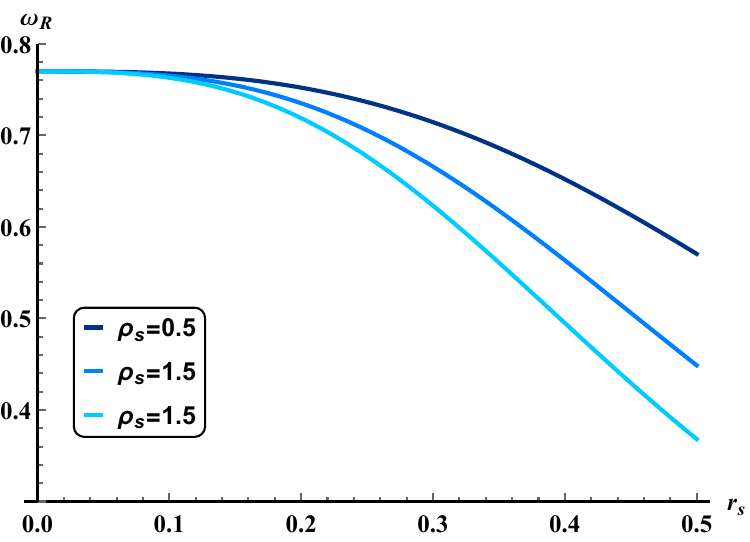} \hspace{-0.2cm}
    \caption{Left panel: shadow radius in terms of $r_s$ for three values of $\rho_s$, taking $M=1$. Left panel: the real part of the QNMs using the eikonal method and considering $M=1$ and $l=4$.}\label{fig:QNM}
\end{figure}

The Klein-Gordon equation considering the Taylor expansion of $Q$ in $r^*$ coordinates has the following eigenvalues, that are known as QNMs
\begin{eqnarray}\label{Qrstar}
\frac{Q(r_c)}{\sqrt{2\dfrac{d^2 Q(r^*)}{d r^{*^2}}\bigg|_{r^*=r_c}}}=i\left(n+\frac{1}{2}\right).
\end{eqnarray}
It should be noted that the above result is valid for asymptotically flat spacetime \cite{cardoso2009geodesic}. By substituting the value of $Q(r_c)$ from Eq. \eqref{Qr} into the previous equation, it reduces to
\begin{eqnarray}
\omega_{\text{QNM}}= l\sqrt{\left(\frac{f(r_c)}{r_c^2}\right)}-i\frac{n+\frac{1}{2}}{\sqrt{2}}\sqrt{\frac{-r_c^2}{f(r_c)}\left(\frac{d^2}{d r^{*^2}}\frac{f(r^*)}{r^{*^2}}\right)_{r=r_c}}\, .
\end{eqnarray}
For an asymptotically flat spacetime, the shadow is calculated from \cite{perlick2022calculating,jha2025thermodynamics}
\begin{eqnarray}
r_{sh}=\frac{r_c}{\sqrt{f(r_c)}} .
\end{eqnarray}
The Lyapunov exponent is a measure of how quickly light rays diverging around a photon's orbit fall into the black hole or escape to infinity. For the metric of the form \eqref{ds2}, the Lyapunov exponent is calculated from
\begin{eqnarray}
|\lambda|=\frac{1}{\sqrt{2}}\sqrt{\frac{-r^2}{f(r)}\left(\frac{d^2}{d r^{*^2}}\frac{f(r^*)}{r^{*^2}}\right)}\bigg|_{r=r_c},
\end{eqnarray}
which determines the instability time scale of unstable null circular geodesics. Using the equation \eqref{r_rstar}, the Lyapunov exponent is obtained from
\begin{eqnarray}
|\lambda|=\frac{1}{\sqrt{2}}\sqrt{\frac{-r^2}{f(r)}
\left(
f(r)\left(
\frac{\partial}{\partial r} f(r)\frac{\partial}{\partial r}+f(r)\frac{\partial^2}{\partial r^2}
\right)
\frac{f(r)}{r^2}
\right)}\bigg|_{r=r_c}\, .
\end{eqnarray}
In the eikonal regime, or in the limit of geometric optics, the Lyapunov exponent of the null geodesics is related to the imaginary part of the QNM in the form
\begin{eqnarray}
\omega_{\text{QNM}}=\frac{1}{r_{sh}} l - i \left(n+\frac{1}{2}\right)|\lambda|.
\end{eqnarray}
Here, $n$ is the overtone number. Figure \ref{fig:QNM} illustrates the variation of the shadow and the real part of the QNM in terms of $r_s$, and it can be seen that an increase in $r_s$ ($\rho_s$) results in an increase in the black hole shadow and a reduction in the real part of the QNMs.

\subsection{Greybody factors}

\begin{figure}[ht!]
  \includegraphics[width=6.5cm]{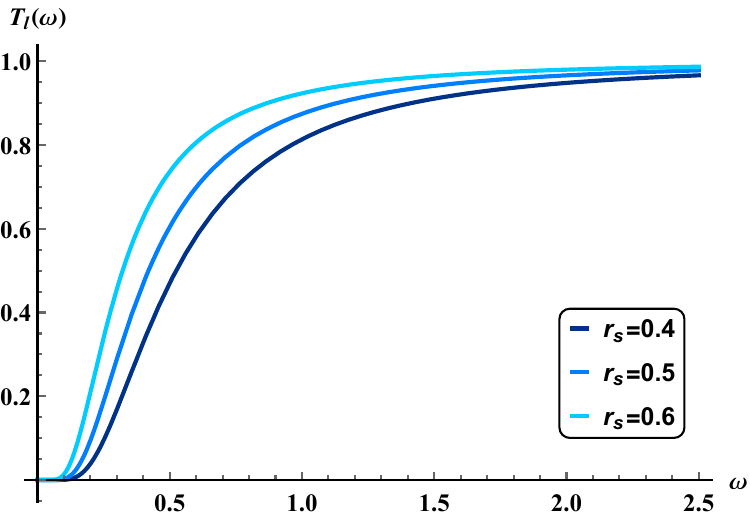} \hspace{0.5cm}
  \includegraphics[width=6.5cm]{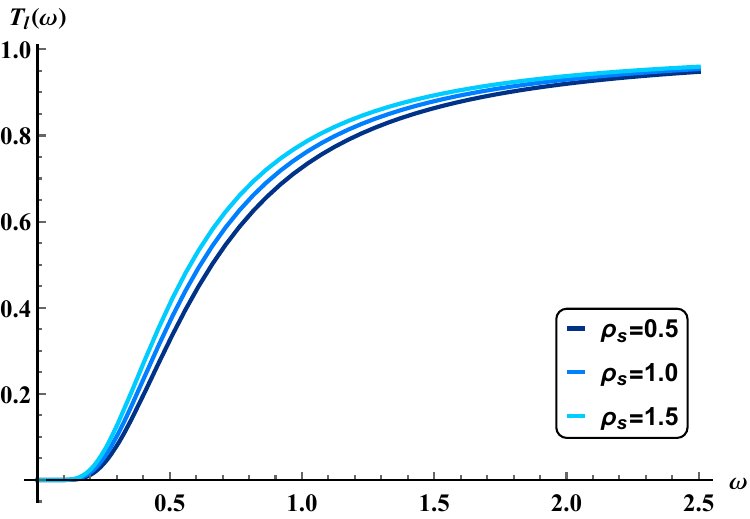} \hspace{-0.2cm}
    \caption{Graybody bounds considering $M=1$, $l=1$ and $s=0$. On the left $\rho_s=1$ and on the right $r_s=0.3$.}\label{fig:GBB}
\end{figure}

Graybody factors represent the modifications in the emission spectrum of black holes due to the presence of potential barriers \cite{kanti2004black,gray2018greybody,harmark2010greybody}. They consider the effects of the black hole's gravitational field on the emitted radiation, its spin, and the dimensionality of spacetime.
These factors modify the blackbody spectrum of Hawking radiation, allowing the properties of the black hole to be identified. \cite{kanti2014greybody,konoplya2020grey}. To calculate the graybody factor, the Regge-Wheeler potential is used to solve the wave equation of the emitted radiation. Using Eq. \eqref{VRW}, an upper bound for the graybody factor is obtained by \cite{boonserm2019greybody,jha2024observational}.
\begin{eqnarray}
T_l(\omega)\geq \text{sech}^2 \left(\frac{1}{2\omega}\int_{r_h}^{\infty}V_{\rm eff}(r) \frac{dr}{f(r)}\right),\label{gbb}
\end{eqnarray}
where $r_h$ refers to the horizon radius calculated in \eqref{horizon} and $\omega$ represents the frequency. Figure \ref{fig:GBB} shows the graybody boundaries in terms of frequency for a given set of initial values.
The graybody factor for perfect blackbody radiation is equal to one, but due to Hawking radiation, this value decreases. It is evident that, holding other parameters constant, a larger $r_s$ ($\rho_s$) corresponds to a larger graybody factor.

\begin{figure}[ht!]
  \includegraphics[width=6.5cm]{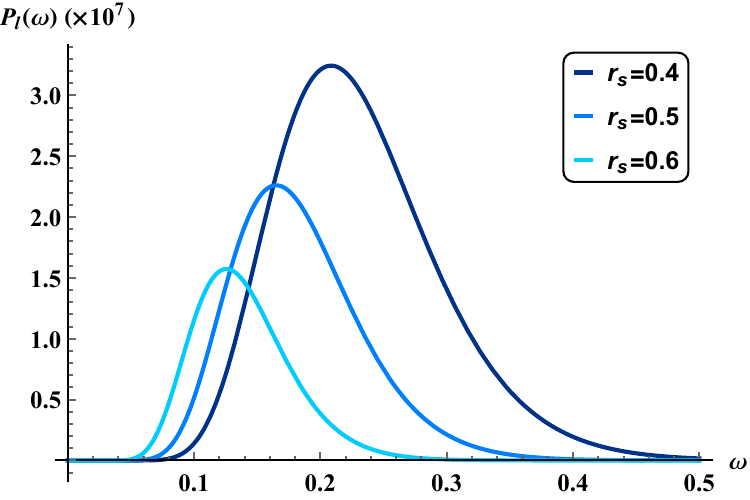} \hspace{0.5cm}
  \includegraphics[width=6.5cm]{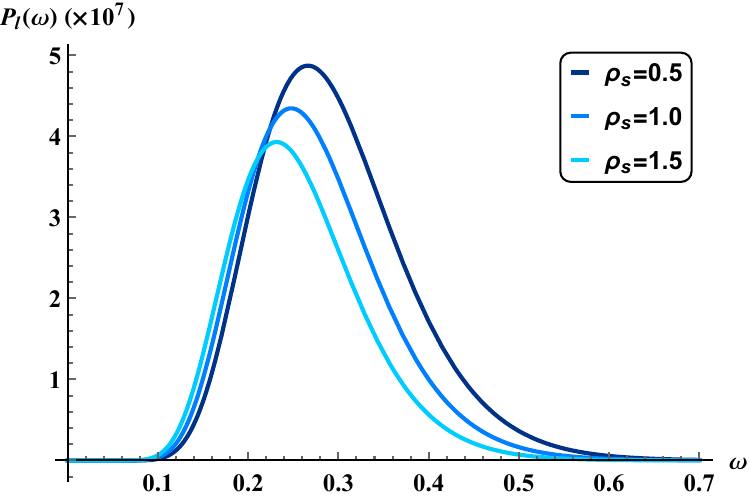} \hspace{-0.2cm}
    \caption{Emitted power for the cases $M=1$, $l=1$ and $s=0$. On the left $\rho_s=1$ and on the right $r_s=0.3$.}\label{fig:GBP}
\end{figure}

The power emitted by a black hole is the rate of energy radiation due to Hawking radiation (\cite{harris2003hawking,kanti2014hawking}). It is quantified as the energy emitted per unit time and is expressed as \cite{miao2017hawking}
\begin{eqnarray}
P_{l}(\omega)=\frac{A}{8\pi^2}T_l(\omega)\frac{\omega^3}{\exp(\omega / T_{\rm H})-1},
\end{eqnarray}
where $T_l(\omega)$ is the graybody factor, $\omega$ is the frequency of the emitted particles, $T_H$ refers to Hawking radiation, and $A$ is the horizon area.
Figure \ref{fig:GBP} illustrates the emitted power for specific initial values. It shows that, with other parameters constant, an increase in $r_s$ $(\rho_s)$ results in a reduction in the emitted power and shifts the peak emitted power to a lower frequency.

\begin{figure}[ht!]
  \includegraphics[width=5.3cm]{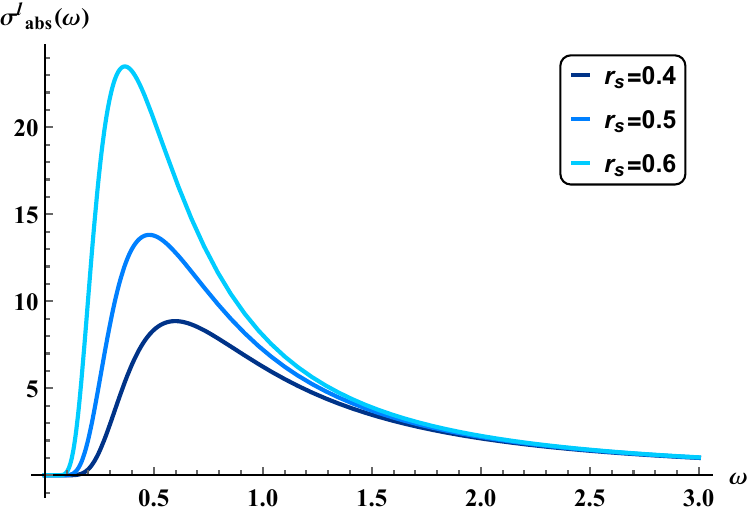} \hspace{0.1cm}
  \includegraphics[width=5.3cm]{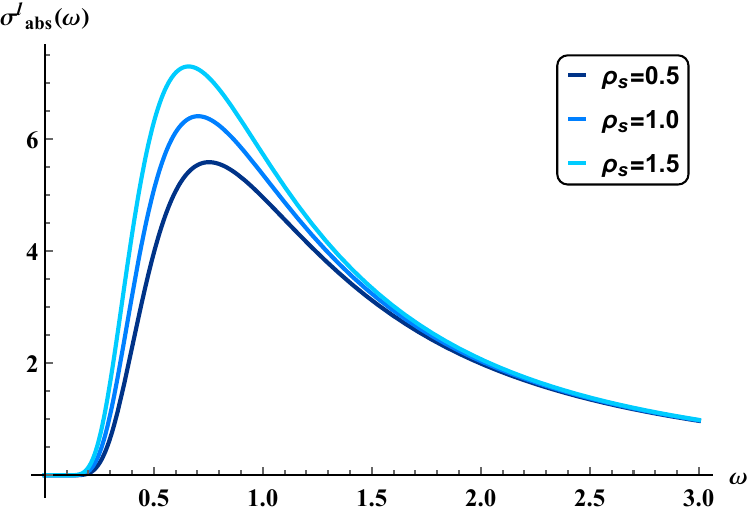} \hspace{0.1cm}
  \includegraphics[width=5.3cm]{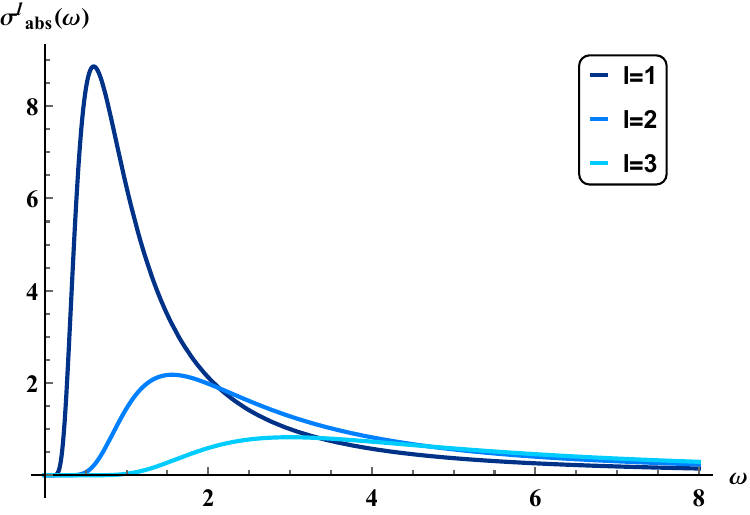} \hspace{-0.2cm}
    \caption{Absorption cross section with parameters $M=1$ and $s=0$. On the left $\rho_s=1$ and $l=1$, in the center $r_s=0.3$ and $l=1$, on the right $\rho_s=1$ and $r_s=0.4$.}\label{fig:GBA}
\end{figure}

The absorption cross section quantifies the probability of a particle being absorbed by a black hole. For the $l$th mode it  is calculated  using \eqref{gbb}  \cite{doran2005fermion,decanini2011universality}
\begin{eqnarray}
\sigma_{abs}^l(\omega) =\frac{\pi (2l+1)}{\omega^2} |T_l(\omega)|^2.
\end{eqnarray}
Figure \ref{fig:GBA} represents the absorption cross-section curves for certain parameters. It is obvious from it that, keeping other parameters constant, an increase in $r_s$ $(\rho_s)$ results in a larger absorption maximum, which shifts to a lower frequency. Furthermore, for a larger value of $l$, the absorption cross-section decreases and its maximum shifts to a higher frequency.

\section{Topological Characteristic}\label{Sec5}

In this section, we study some properties of black holes from a topological perspective. To do so, we define a potential based on the investigated property, to which we assign vectors in spherical coordinates $\phi_r^i$ and $\phi_\theta^i$, normalized in the form $n_r^i=\phi_r^i/||\phi||$ and $n_\theta^i=\phi_\theta^i/||\phi||$, respectively \cite{s2024effective,malik2024thermodynamic,hosseinifar2025quasinormal,liu2024light,fang2023revisiting,sadeghi2023bardeen,wu2024thermodynamical,wu2025novel}. 
Using these normalized vectors, the vector space of this potential can be visualized. There may be points in it where the vectors converge or diverge; these coordinates are the zero points of the vector space and can be considered topological defects \cite{fan2023topological,wei2024universal,chen2024thermodynamical,wu2023topological}.
A topological charge, known as the winding number, can be assigned to each point in this space. This charge is calculated by examining the rotation of vectors around each point \cite{wu2023topological,mehmood2023thermodynamic,wang2024thermodynamic}. To do this, a closed contour $c_i$ can be drawn around each point $(\theta,\, r_i)$ in this vector space. Using a variable transformation such as $r=a \cos\vartheta +r_i$ and $\theta=b \sin\vartheta+\frac{\pi}{2}$, the topological charge within each contour can be calculated from \cite{liu2023topological,mehmood2024davies,yasir2024topological}
\begin{eqnarray}\label{winding}
\Omega=w_i=\frac{1}{2\pi}\oint_{c_i} d(\arctan\frac{n_\theta}{n_r}).
\end{eqnarray}
If there are no zero points within this contour, the topological charge will be zero. However, if there is a zero point, rotating the vectors can indicate whether the topological charge is $+1$ or $-1$. \cite{du2023topological,wei2023topology,yerra2022topology}.
Furthermore, the direction of the vectors around a zero point can be used to investigate the topological charge of the zero point, where the clockwise direction refers to a topological charge of $-1$ and the counterclockwise direction of the curve $\phi_\theta-\phi_r$ indicates a topological charge of $+1$ \cite{rizwan2025universal}.
Based on the potential considered, conclusions can be drawn about it. In the following sections, three different potentials will be studied from a topological perspective.

\subsection{Photon Sphere}\label{Sec51}

\begin{figure}[ht!]
  \includegraphics[width=6.8cm]{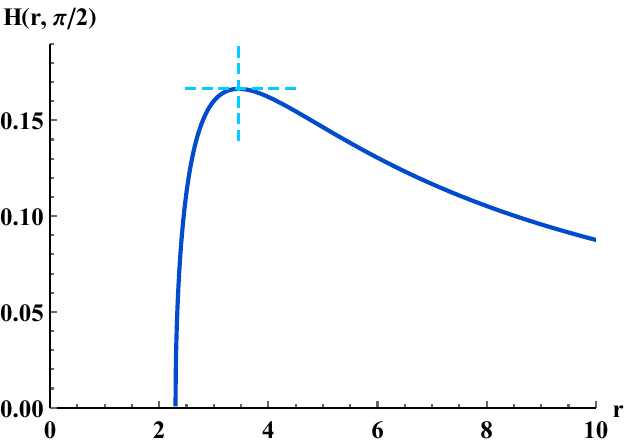} \hspace{-0.2cm}
    \caption{The potential $H(r,\pi/2)$ as a function of $r$. Its extreme is at $\partial_r H\big|_{r=r_c}=0$.}\label{fig:TopoH}
\end{figure}

In Ref. \cite{jha2025thermodynamics}, the photon sphere and the shadow of the black hole were studied. In this section, to investigate the stability or instability of the photon sphere, a potential of the form \cite{alipour2024weak,gashti2024thermodynamic,pantig2025multimodal}
\begin{eqnarray}\label{HPot}
H(r,\theta)=\frac{1}{\sin\theta}\sqrt{\frac{f(r)}{h(r)}},
\end{eqnarray} 
is defined, which can be represented in the vector space using the vectors \cite{sekhmani2024thermodynamic,shahzad5033870topological,wei2020topological}
\begin{align}
\phi_{r}^H =\sqrt{f(r)} \partial_r H(r,\theta),\qquad
\phi_{\theta}^H =\frac{1}{\sqrt{h(r)}}\partial_{\theta} H(r,\theta).
\end{align}
Figure \ref{fig:TopoH} shows the potential $H$ at $\theta=\pi/2$ for $M=1$, $r_s=0.3$ and $\rho_s=1$, which allows us to appreciate the existence of an extreme where the radial derivative of $H$ becomes zero.

\begin{figure}[ht!]
  \includegraphics[width=5.7cm]{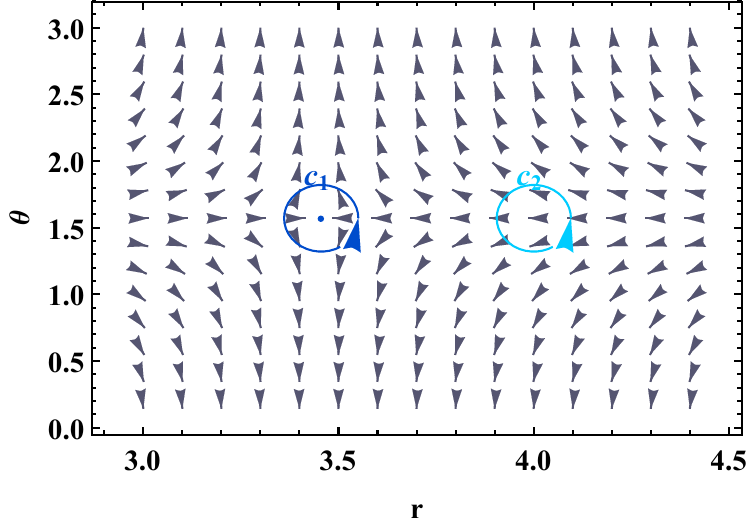} \hspace{0.1cm}
  \includegraphics[width=4.4cm]{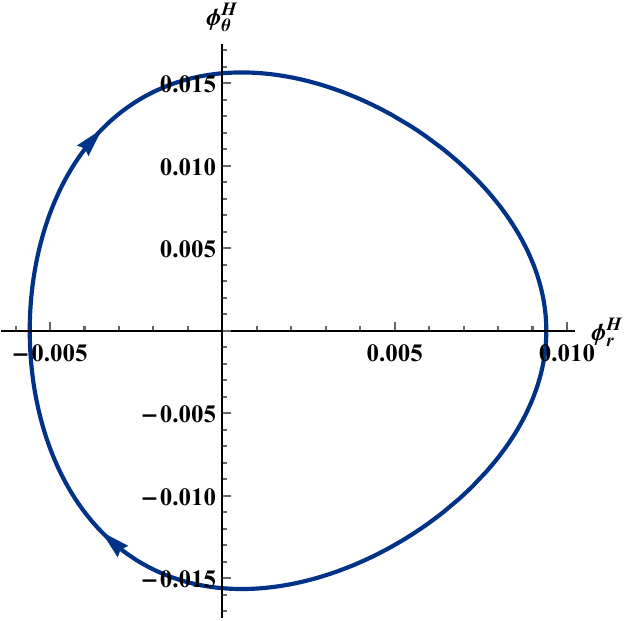} \hspace{0.1cm}
  \includegraphics[width=5.8cm]{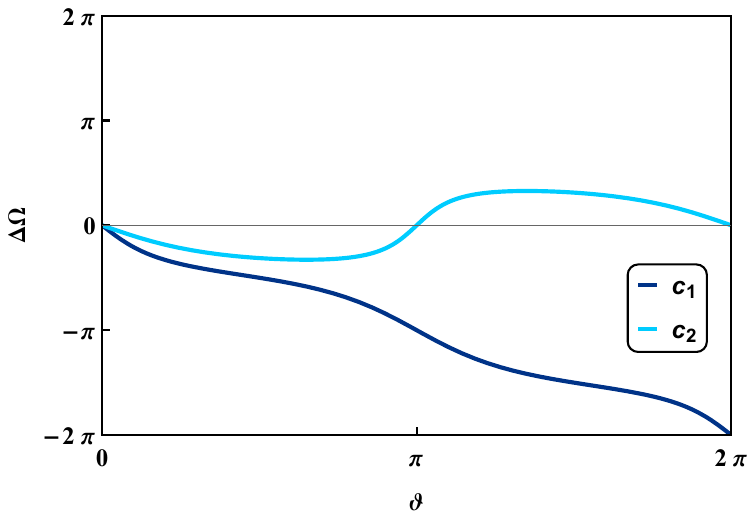} \hspace{-0.1cm}\\
    \caption{
Left panel: Vector space of the potential $H$ with closed contours $c_1$ and $c_2$ drawn around $r_1=3.45582$ and $r_2=4.0$, respectively.
Middle panel: The behavior of $\phi\theta-\phi r$ is shown for the zero point $r_1$. The clockwise direction indicates that $\omega_1=1$.
Right panel: Variation of $\Omega$ in terms of $\vartheta$, with $a=b=0.3$. The topological charge inside the closed boundary $c_1$ is $w_1=-1$, while the topological charge inside $c_2$ is $\omega_2=0$. The relevant parameters have been chosen to be $M=1$, $r_s=0.3$ and $\rho_s=1$.}\label{fig:TopoPS}
\end{figure}

In Fig. \ref{fig:TopoPS}, the normalized vector space representation of these vectors is shown for the case $M=1$, $r_s=0.3$ and $\rho_s=1$.
Two closed contours, $c_1$ and $c_2$, are drawn around the zero point $r_1=3.45582$ and an arbitrary point $r_2=4.0$, respectively. Considering the rotation direction of the $\phi_r-\phi_\theta$ curve in the middle panel, as well as the variations of $\Delta\Omega$ in the right panel of Fig. \ref{fig:TopoPS}, we conclude that the closed contour $c_2$ contains no topological charge, while the topological charge inside the contour $c_1$ is $-1$. This charge indicates the instability of the photon sphere located at $r_1$ \cite{wei2020topological}.

\subsection{Temperature}\label{Sec52}

\begin{figure}[ht!]
  \includegraphics[width=5.7cm]{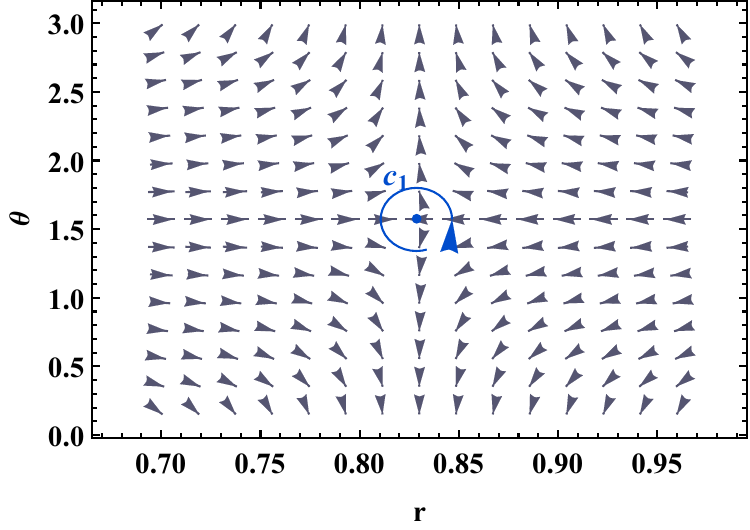} \hspace{0.1cm}
  \includegraphics[width=4.4cm]{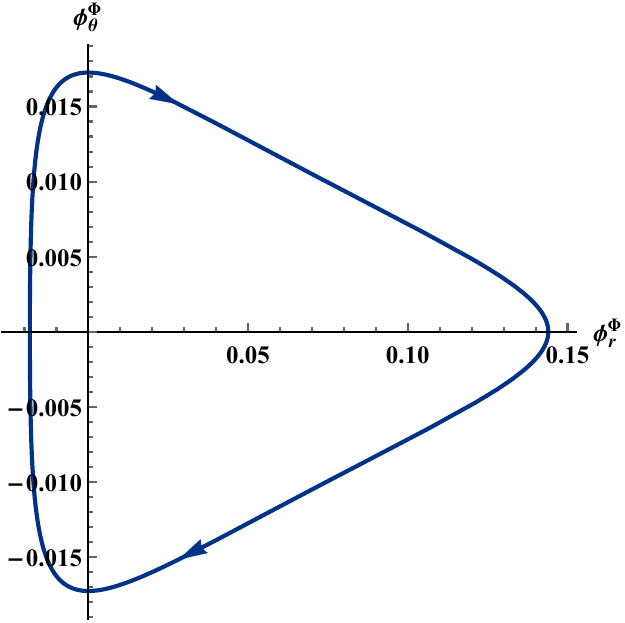} \hspace{0.1cm}
  \includegraphics[width=5.8cm]{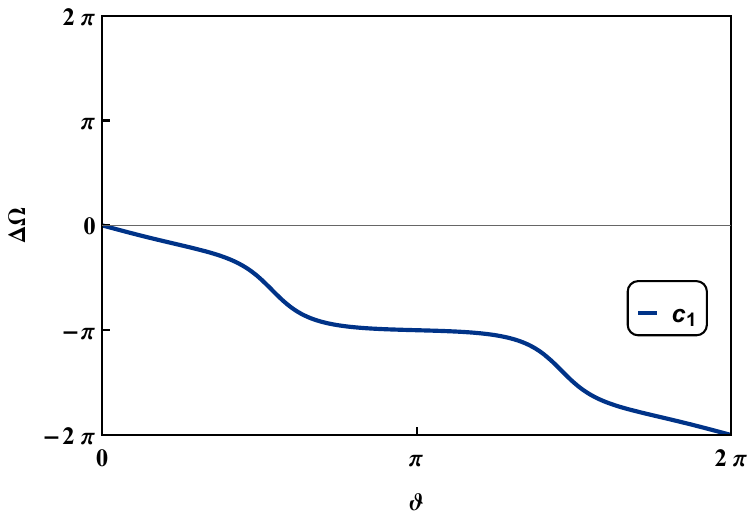} \hspace{-0.1cm}\\
    \caption{
Left panel: Vector space of the potential $\Phi$ containing a zero point at $r_1=0.82854$.
Middle panel: Curve $\phi_\theta-\phi_r$ around zero point $r_1$. Clockwise direction means that $\omega_1=1$.
Right panel: Changes in $\Omega$ as a function of $\vartheta$ for the closed contour $c_1$ ($a=b=0.3$), indicating a topological charge of $-1$. The parameters are chosen as $r_s=0.5$ and $\rho_s=1$.}\label{fig:TopoTemp}
\end{figure}

Another feature of black holes that can be examined from a topological perspective is the black hole Hawking temperature \cite{robson2019topological,robson2019hawking}. In section \ref{Sec4} we studied the black hole Hawking temperature and found that for $2 \sqrt{\pi \rho_s} r_s > 1$, a phase transition to the black hole Hawking temperature occurs. In this section, this transition will be examined using a topological approach.
For this purpose, a thermodynamic potential is defined as \cite{yerra2023topology,chen2023thermodynamics,bhattacharya2024topological}
\begin{eqnarray}
\Phi =\frac{1}{\sin \theta} T_H,
\end{eqnarray}
which can be represented in the vector field defined by the following vectors \cite{jeon2025stability,chen2024thermodynamic,ali2024topology}
\begin{equation}
\phi^{\Phi}_r=\partial_{r_h}\Phi,\qquad
\phi^{\Phi}_\theta=\partial_\theta\Phi.
\end{equation}
In Fig. \ref{fig:TopoTemp}, the vector space of potential $\Phi$ is illustrated for $r_s=0.5$ and $\rho_s=1$, and the zero point $r_1=0.82854$ can be seen.
To analyze the topological charge of the zero point $r_1=0.82854$, we draw a closed contour $c_1$ around it and, choosing $a=b=0.3$, calculate the topological charge using the equation \eqref{winding}. The variations of $\Omega$ with respect to $\vartheta$ for the zero point $r_1$ on the closed contour $c_1$ are shown in figure \ref{fig:TopoTemp}. It is evident that this point has a topological charge of $-1$. This result is clearly seen in the central panel, considering the clockwise rotation of the $\phi_r-\phi_\theta$ curve.
Therefore, we can conclude that by choosing $2 \sqrt{\pi \rho_s} r_s>1$, a phase transition occurs at the critical point $r_c$ in the Hawking temperature of the black hole, and this critical point is conventional \cite{wei2022topology}.

\subsection{Generalized Free Energy}\label{Sec53}
\begin{figure}[ht!]
  \includegraphics[width=6.5cm]{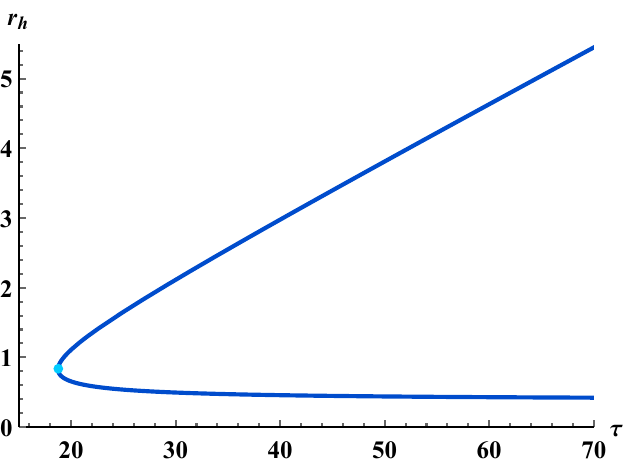} \hspace{-0.2cm}\\
    \caption{$r_h-\tau$ curve \eqref{xtr} for the values $r_s=0.5$ and $\rho_s=1$. At $\tau=18.75922$ the sign of the slope of the curve changes.}\label{fig:TopoTau}
\end{figure}

The last property we will study from a topological perspective is the generalized free energy of the black hole outside the shell surrounding the horizon. This energy is obtained from \cite{gashti2024topology,di2024topological,hung2023topology}
\begin{equation}
\mathcal{F}=M(r_h)-\frac{S}{\tau},
\end{equation}
where $M(r_h)$ is the mass of the black hole as calculated in the equation \eqref{Mh}, $\tau$ is a parameter representing the inverse temperature of the black hole, and $S$ denotes the entropy of the black hole, which using the equations \eqref{Mh} and \eqref{TH} is defined as
\begin{eqnarray}
S=\int\frac{d M(r_h)}{T_H} dr_h=\pi r_h^2,
\end{eqnarray}
and is similar to the Schwarzschild black hole entropy. The relevant vector space associated to the potential $\mathcal{F}$ is defined as \cite{rizwan2023topological,eslam2024thermodynamic,fairoos2023topological,Wei2022Black}
\begin{equation}
\phi_r^{\mathcal{F}}=\partial_{r_h}\mathcal{F},\qquad
\phi_\theta^{\mathcal{F}}=-\cot \theta \csc \theta.
\end{equation}
The zero points of this potential are located at $(r_c,\,\pi/2)$, where $r_c$ is computed from $\partial_{r_h}\mathcal{F}\big|_{r_h=r_c}=0$. Therefore, by setting the radial derivative $\mathcal{F}$ to zero, we can find a relationship between $\tau$ and $r_h$, given by
\begin{equation}\label{xtr}
\tau=\frac{4 \pi  r (r+r_s)^2}{(r+r_s)^2-4 \pi  \rho_s r_s^4}\Bigg|_{r_h=r_c} .
\end{equation}
Figure \ref{fig:TopoTau} shows the $r_h-\tau$ curve obtained from \eqref{xtr} for the values $r_s=0.5$ and $\rho_s=1$.
The sign of the slope of this curve changes at $(0.82854,\,18.75922)$, where it becomes infinite, with two branches appearing for $\tau>18.75922$.

\begin{figure}[ht!]
  \includegraphics[width=5.7cm]{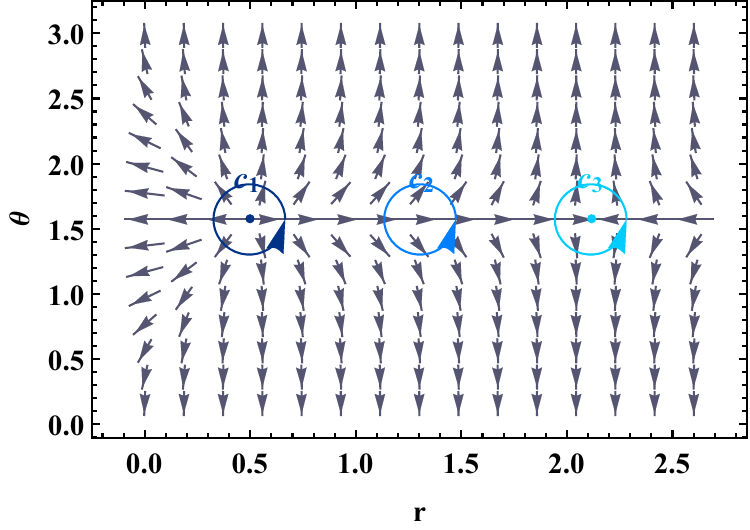} \hspace{0.1cm}
  \includegraphics[width=4.4cm]{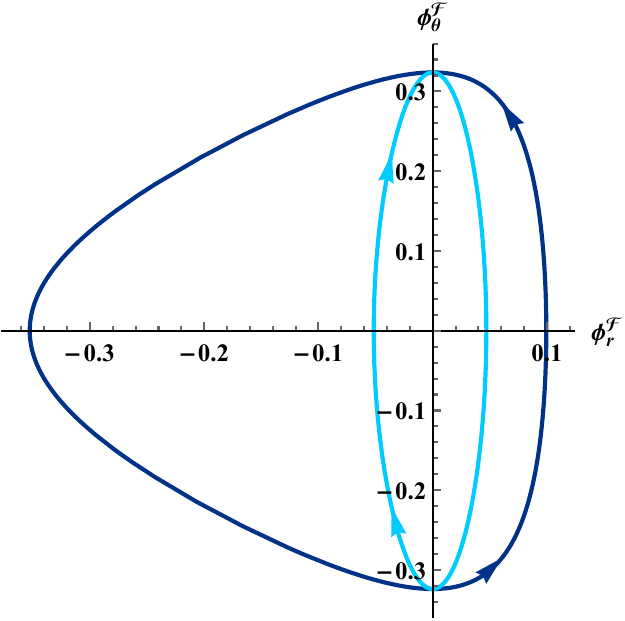} \hspace{0.1cm}
  \includegraphics[width=5.8cm]{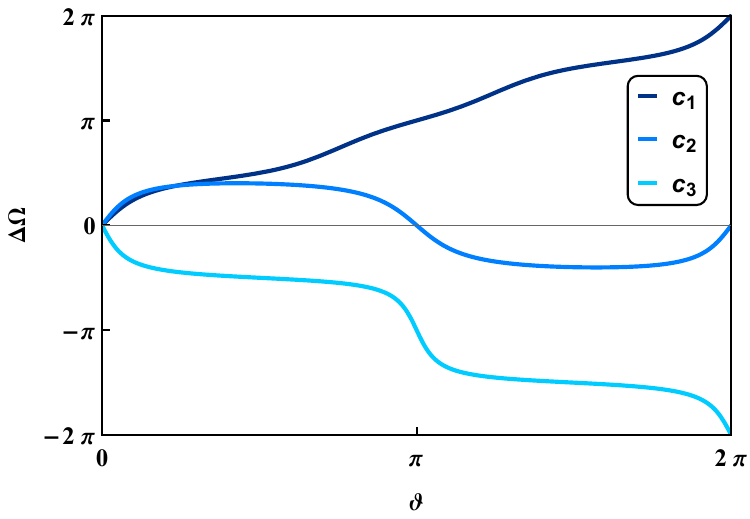} \hspace{-0.1cm}\\
    \caption{
Left panel: Vector space of the potential $\mathcal{F}$. The zero points $r_1=0.49556$ and $r_3=2.11263$ are indicated, as well as an arbitrary point at $r_2=1.30409$.
Middle panel: $\phi_\theta$ versus $\phi_r$ for zero points $r_1$ and $r_2$. The direction of the curves indicates that $\omega_1=1$ and $\omega_3=-1$.
Right panel: The variation of $\Omega$ with respect to $\vartheta$ under the assumption $a=b=0.3$ is shown for the closed contours $c_1$, $c_2$ and $c_3$, which enclose the points $r_1$, $r_2$ and $r_3$, respectively. The corresponding topological charge of these contours is given by $\omega_1=1$, $\omega_2=0$ and $\omega_3=-1$.
The parameters are set as $r_s=0.5$, $\rho_s=1$ and $\tau=30$.
}\label{fig:TopoHelm}
\end{figure}

The first graph in Figure \ref{fig:TopoHelm} shows the vector space associated with the potential $\mathcal{F}$ for the parameters $r_s=0.5$, $\rho_s=1$ and $\tau=30$. The two zero points $r_1=0.49556$ and $r_3=2.11263$ are highlighted in this space, along with the arbitrary point $r_2=1.30409$.
To calculate the topological charge of these points, closed contours $c_1$, $c_2$, and $c_3$ have been drawn around the points $r_1$, $r_2$, and $r_3$, respectively. By choosing $a=b=0.3$, we can calculate the integral from \eqref{winding} for the specified points.
In the middle panel of Fig. \ref{fig:TopoHelm} we plot the curve $\phi_{\theta}-\phi_{r}$  to show the zero point $r_h$.
The last graph in Fig. \ref{fig:TopoHelm} shows the variation of $\Omega$ with respect to $\vartheta$ for the three contours $c_1$, $c_2$ and $c_3$. As indicated, these contours correspond to topological charges of $\omega_1=+1$, $\omega_2=0$ and $\omega_3=-1$, respectively. This result is also evident from the middle panel of Fig. \ref{fig:TopoHelm}, taking into account the rotation direction of the $\phi_r-\phi_\theta$ curves.
Therefore, the vector space of the potential $\mathcal{F}$ contains two zero points, and the total topological charge is zero. This means that, from a topological classification perspective, this black hole falls into the same class as the Reissner-Nordstr{\"o}m black hole \cite{Wei2022Black}.

\section{Summary and conclusions}\label{Sec6}

Considering a Schwarzschild black hole immersed in a Hernquist dark matter halo (SBH-HDM), the mass of the system and the temperature at the horizon are determined. In some special cases of dark matter, there is some remnant mass, whereas in the absence of dark matter, when the horizon radius tends to zero, the horizon temperature tends to infinity. This is why many scientists are interested in investigating modified gravity to obtain remnant mass. The importance of remnant mass lies in the process of black hole evaporation; at the end of this process, a special value of the mass is obtained, called the remnant mass, and in this case, the information is not lost.
Our findings revealed that, for specific choices of $r_s$ and $\rho_s$, a phase transition occurs at the black hole's Hawking temperature, resulting in the emergence of a remnant mass. We established a criterion for parameter selection that facilitates the observation of both the remnant mass and the phase transition at the Hawking temperature.
Next, we explore the influence of the Hernquist-type dark matter distribution on the matter in the thin accretion disk, quasi-normal modes in the eikonal limit, gray-body boundaries, and the thermodynamic topologies of black holes. Initially, we investigate the motion of test particles, circular orbits, and radiative properties of the thin accretion disk around a black hole within a Hernquist-type dark matter halo. We observe that, holding one parameter of the dark matter halo constant, an increase in the other leads to a reduction in energy $E$, while the angular momentum $L$ and angular velocity $\Omega$ increase with respect to the corresponding values in the Schwarzschild black hole case.
 Moreover, at smaller radii, the impact of the dark matter model parameters $r_{s}$ and $\rho_{s}$ decreases significantly.
In the equatorial plane, we calculated the ISCO radius and found that an increase in either $r_{s}$ or $\rho_{s}$ results in an expansion of this radius.
We find that, for small $r_{s}$, the radiative efficiency increases with rising halo density $\rho_{s}$, while for larger $r_{s}$, this trend is reversed and the efficiency decreases as $\rho_{s}$ increases. Furthermore, we observed that an increase in the parameter $\rho_s$ leads to a shift in this trend at a smaller $r_s$.
Next, we analyze the thin accretion disk in the SBH-HDM scenario as the primary instrument for studying the surrounding spacetime geometry. Using the steady-state Novikov-Thorne model, we numerically calculate key disk properties, including the energy flux and temperature distribution, the differential luminosity, and the spectral luminosity, for geometrically thin accretion disks. These results were subsequently compared with those obtained for a Schwarzschild black hole in the framework of general relativity.

Our analysis revealed that the Schwarzschild black hole exhibits a higher energy flux, while an increase in $r_s$ causes this value to decrease and shifts its peak to a larger radius. Furthermore, we examined the radiation temperature and observed a decrease compared to the Schwarzschild black hole. In our study of the differential luminosity, we observed a significant increase with increasing $r_s$, and the maximum now occurs at a larger radius.

We then delve deeper into the black hole's quasi-normal modes, investigating the interaction between them and the black hole shadow within the eikonal regime. 
We have shown that, in the eikonal regime, the Lyapunov exponent of the null geodesics is related to the imaginary part of the QNM, which can illustrate the stability or instability of the system. Furthermore, the shadow can be obtained by taking the inverse of the real part of the QNM, which is comparable with the other methods.
In particular, we find that holding other parameters constant while increasing $r_s$ leads to an expansion of the black hole shadow and a reduction in the real part of the quasi-normal modes. We then examine the scattering surrounding the black hole, analyzing the graybody factor, the emitted power, and the absorption cross section. Our results indicated that an increase in $r_s$ ($\rho_s$), with other parameters constant, enhances both the graybody factor and the absorption, shifting the absorption peak to lower frequencies.
On the contrary, this increase results in a decrease in the emitted power, shifting its peak toward lower frequencies. In the final segment, we investigate the black hole's photon sphere from a topological perspective, revealing that this black hole possesses an unstable photon sphere, which contributes to the formation of its shadow. Finally, we examine the black hole's thermodynamic potentials from a topological perspective.
We explore the phase transition at the Hawking temperature for $2 \sqrt{\pi \rho_s} r_s>1$ and identify this critical point as conventional. Furthermore, we find that the generalized out-of-shell free energy of the black hole exhibits two distinct phase transitions, while the topological charge in the vector space of this thermodynamic potential is zero. This allows us to precisely classify the topological nature of this black hole, placing it within the realm of Reissner-Nordstr{\"o}m black holes.
Extending the present analysis to the class of new rotating spacetimes (considered astrophysically more realistic than those admitting a smooth and asymptotically flat expansion) represents a compelling direction for future research, in which we are currently making rapid progress.

\section*{Acknowledgements}
The research of L.M.N., S.Z. and H.H. was supported by the Q-CAYLE project, funded by the European Union-Next Generation UE/MICIU/Plan de Recuperacion, Transformacion y Resiliencia/Junta de Castilla y Leon (PRTRC17.11), and also by project PID2023-148409NB-I00, funded by MICIU/AEI/10.13039/501100011033. Financial support of the Department of Education of the Junta de Castilla y Leon and FEDER Funds is also gratefully acknowledged (Reference: CLU-2023-1-05). Additionally, H. H. is grateful to Excellence project FoS UHK 2203/2025-2026 for the financial support.

\bibliography{ref.bib}


\end{document}